\documentclass[amsmath,amssymb,twocolumn,superscriptaddress]{revtex4-2}

\usepackage{graphicx} 
\usepackage{nicefrac}
\usepackage{dcolumn}
\usepackage{bm}
\usepackage{array}
\usepackage{chngcntr}
\usepackage{amssymb}
\usepackage{amsmath}
\usepackage{mathtools}
\usepackage{hyperref}
\usepackage{verbatim}
\usepackage{mleftright}
\usepackage{sidecap}

\usepackage[colorinlistoftodos]{todonotes}

\counterwithin*{section}{part}

\global\long\def\cvec2#1#2{\left(\begin{array}{c}
#1\\
#2
\end{array}\right)}
\global\long\def\mat2#1#2#3#4{\left[\begin{array}{cc}
#1 & #2 \\
#3 & #4
\end{array}\right]}

\global\long\def\re{\mathrm{Re}}%
\global\long\def\im{\mathrm{Im}}%
\global\long\def\xnorm{\mathcal{N}}%
\global\long\def\xr{\mathcal{R}}
\global\long\def\xb{\mathcal{B}}

\global\long\def\e{\mathrm{\mathbb{E}}}%
\global\long\def\eO{\mathrm{\mathbb{E}}_{\bar{\bm O}}}%

\global\long\def\norm#1{\left\Vert #1\right\Vert }%
\global\long\def\trace#1{\underset{#1}{\mathrm{Tr}}}%

\global\long\def\rm#1{\mathrm{#1}}%

\global\long\def\xlim#1#2{\lim_{#1\rightarrow#2}}%

\global\long\def\bb#1{\bm{#1}}%
\global\long\def\mb#1{\bm{#1}}

\global\long\def\ex#1{e^{#1}}%
\global\long\def\lr#1{\mleft(#1\mright)}
\global\long\def\lrb#1{\mleft[#1\mright]}

\global\long\def\E{\mathbb{E}}%
\global\long\def\cZ{\mathcal{Z}}%
\global\long\def\cF{\mathcal{F}}%

\global\long\def\S{\mathbb{S}}%
\global\long\def\O{\mathbb{O}}%
\global\long\def\P{\mathbb{P}}%

\begin{document}

\title{Singular vectors of sums of rectangular random matrices and \\ optimal estimation of high-rank signals: the extensive spike model}

\author{Itamar D. Landau}
\affiliation{Dept. of Applied Physics, Stanford University, Stanford, CA 94305, USA}
\email{idlandau@stanford.edu}
\author{Gabriel C. Mel}
\affiliation{Neuroscience Graduate Program, Stanford University, Stanford, CA 94305, USA}
\author{Surya Ganguli}
\affiliation{Dept. of Applied Physics, Stanford University, Stanford, CA 94305, USA}

\date{\today}

\begin{abstract}

Across many disciplines spanning from neuroscience and genomics to machine learning, atmospheric science and finance, the problems of denoising large data matrices to recover hidden signals obscured by noise, and of estimating the structure of these signals, is of fundamental importance. A key to solving these problems lies in understanding how the singular value structure of a signal is deformed by noise. This question has been thoroughly studied in the well-known spiked matrix model, in which data matrices originate from low-rank signal matrices perturbed by additive noise matrices, in an asymptotic limit where matrix size tends to infinity but the signal rank remains finite. We first show, strikingly, that the singular value structure of large finite matrices (of size $\sim1000$) with even moderate-rank signals, as low as $10$, is not accurately predicted by the finite-rank theory, thereby limiting the application of this theory to real data. To address these deficiencies, we analytically compute how the singular values and vectors of an arbitrary {\it high}-rank signal matrix are deformed by additive noise. We focus on an  asymptotic limit corresponding to an $\textit{extensive}$ spike model, in which {\it both} the signal rank and the size of the data matrix tend to infinity at a constant ratio. We map out the phase diagram of the singular value structure of the extensive spike model as a joint function of signal strength and rank. We further exploit these analytics to derive optimal rotationally invariant denoisers to recover the hidden $\textit{high}$-rank signal from the data, as well as optimal invariant estimators of the signal covariance structure. Our extensive-rank results yield several conceptual differences compared to the finite-rank case: (1) as signal strength increases, the singular value spectrum does not directly transition from a unimodal bulk phase to a disconnected phase, but instead there is a new bimodal connected regime separating them (2) the signal singular vectors can be partially estimated $\textit{even}$ in the unimodal bulk regime, and thus the transitions in the data singular value spectrum do not coincide with a detectability threshold for the signal singular vectors, unlike in the finite-rank theory; (3) signal singular values interact nontrivially to generate data singular values in the extensive-rank model, whereas they are non-interacting in the finite-rank theory; (4) as a result, the more sophisticated data denoisers and signal covariance estimators we derive, that take into account these nontrivial extensive-rank interactions, significantly outperform their simpler, non-interacting, finite-rank counterparts, even on data matrices of only moderate rank. Overall, our results provide fundamental theory governing how high-dimensional signals are deformed by additive noise, together with practical formulas for optimal denoising and covariance estimation.

\end{abstract}

\maketitle

\tableofcontents

\section{Introduction}
Estimating structure in high-dimensional data from noisy observations constitutes a fundamental problem across many disciplines, especially in the age of {\it big-data}. A common scenario is that such data are presented as a large matrix. Such matrices could contain, for example, the observed time series of many recorded neurons in neuroscience, the expression level of many genes across many conditions in genomics, or the time series of many stock prices in finance. Given such data matrices, one often wishes to: (1) understand the structure of the data via its singular value decomposition; (2) denoise the data in order to find clean signals hidden in the data, and (3) estimate the covariance structure of these clean hidden signals. These hidden signals could correspond for example to temporally correlated cell assemblies in neuroscience, gene modules in genomics, or sectors of correlated stocks in finance. 

These three problems of data understanding, data denoising, and signal-covariance estimation, raise fundamental new challenges in the era of big-data, where the number of observations (i.e. the length of time series, or the number of conditions) is often comparable to the number of variables (i.e. the number of recorded neurons, genes, or stock prices). As a result, tools from random matrix theory (RMT) designed for this high-dimensional regime have grown in prominence across a wide range of disciplines including neuroscience \cite{Rumyantsev2020-nl}, psychology \cite{Saxe2019-eg}, genetics \cite{luo2006genetics}, finance \cite{Plerou2002}, machine learning \cite{Pennington2017,Pennington2017-za, Pennington2018-fy, Lampinen2018-sl,martin2021implicit, Wei2022}, atmospheric science \cite{santhanam2001atmospheric,SANTOS2020_atmospheric}, wireless communications \cite{Tulino2004}, integrated energy systems \cite{zhu2019energy_systems}, and magnetic resonance imaging (including spectroscopy \cite{Mosso2022-at, Clarke2022-rj}, diffusion \cite{Tax2022-fj}, and functional-MRI \cite{Bansal2021,Zhu2022}).

In this work, we develop new RMT tools in order to quantitatively study the basic question of how the singular value decomposition (SVD) of an arbitrary high-dimensional hidden signal matrix is deformed under additive observation noise. Based on this understanding of the relation between the data and signal SVDs, we go on to derive both optimal denoisers of the data to recover the hidden signal, as well as optimal estimators for the signal covariance.

An influential line of prior related research has studied spiked matrix models, focusing on an asymptotic limit in which the size of the hidden signal matrix tends to infinity but its rank remains finite \cite{Baik2005,Loubaton2011,Benaych-Georges2012,Shabalin2013,el-alaoui18a,Barbier2020,Aubin2021,Ke2021}. These works consider the addition of a random noise matrix to the signal to generate a data matrix, and they study how the singular values and singular vectors of the data are related to those of the signal. The finite number of signal eigenvalues or singular values constitute a set of ``spikes'' in the signal spectrum, hence the name spiked matrix model.

A key observation in these models is that the addition of noise to the signal yields a data matrix that: (1) has inflated singular values relative to the signal, and (2) has singular vectors that are rotated relative to the signal. For the finite-rank rectangular spiked matrix model, both the degree of singular-value inflation and the angle of the singular vector rotation can be explicitly computed \cite{Benaych-Georges2012, Shabalin2013}. Notably, in this finite-rank regime, the multiple spikes do not interact as they get deformed from signal to data. This means that to predict the mapping from a given signal singular value to the corresponding data singular value, as well as the angle between a data and signal singular vector, one only needs to know the noise distribution and the singular value of the signal spike in question; one does not need to know all the other signal singular values. This underlying simplicity in the relation between signal and data spectral structure implies that one can optimally denoise data, and optimally estimate signal covariance, by applying shrinkage functions that act independently, albeit non-linearly, on each singular value or eigenvalue of the data \cite{Shabalin2013,Gavish2014,Gavish2017}. The idea is that these shrinkage functions that shrink data singular values independently, partially reverse the independent singular-value inflation and compensate for the independent singular-vector rotation, both due to additive noise. 

In this work, however, we demonstrate that the assumptions and consequences of the finite-rank model may constitute a significant limitation for the practical application of this theory and its associated estimation techniques. For example, below we will see that the spectral structure of random matrices of large size (e.g., $1000\times500$), and of even moderate rank (e.g., $10$) cannot be accurately modeled by the finite-rank spiked matrix model. 

This lack of numerical accuracy of the finite-rank theory for large but finite-size matrices of moderate rank could have a significant impact on the three problems of spectral understanding, data denoising, and signal covariance estimation across the empirical sciences, where the effective rank of signals is expected to vary significantly, and sometimes even be quite high. Therefore, it is imperative to develop a new theory that more accurately describes data containing higher-rank signals. We develop that theory by generalizing the finite-rank theory to an {\it extensive-rank} theory in which the rank of the signal matrix is proportional to the size of the signal and data matrices, working in an asymptotic limit where {\it both} the size and rank approach infinity. 

We note that it is not immediately obvious how to extend existing finite-rank results to the extensive regime. The finite-rank theory \cite{Baik2005, Loubaton2011, Benaych-Georges2012} makes use of algebraic formulas for matrices with low-rank perturbations that do not directly generalize, and so one must resort to more elaborate tools from RMT and free probability. Along these lines, powerful theoretical methods have been developed in recent years for studying the eigen-decomposition of sums of square Hermitian matrices \cite{Allez2014}, and deriving techniques for optimally estimating arbitrary square-symmetric matrices from noisy observations \cite{Ledoit2011, Ledoit2012, Bun2016,Bun2017,Ledoit2020,potters_bouchaud_2020}.

However the situation for rectangular matrices, relevant to data from many fields including neuroscience, genomics and finance, lags behind that of square matrices. While the singular value spectrum of sums of rectangular matrices has been calculated \cite{Benaych-Georges2009, Speicher2011, Benaych-Georges2011, Mingo2017}, and a few works have studied optimal denoising of rectangular matrices under a known (usually Gaussian) prior \cite{Troiani2022, Barbier2022, Maillard2022}, there are currently no methods for determining the deformation of the singular vectors of a rectangular signal matrix due to an additive noise matrix. 

The outline of our paper is as follows. In Section \ref{sec:Spiked Matrix Model} we motivate our work with an illustrative numerical study of the spiked matrix model, showing that the finite-rank theory fails to accurately predict the outlier singular values and singular vector deformations in data matrices containing even moderate-rank signals. In section \ref{sec:mathematical prelims} we introduce tools from RMT that we will need to derive our results, including Hermitianization, block matrix resolvents, block Stieltjes transforms and their inversion formulae, and block R-transforms. In Section \ref{sec:SVD of Sums} we study how the singular values and singular vectors of an arbitrary rectangular signal matrix are deformed under the addition of a noise matrix to generate a data matrix. To do so, we derive a subordination relation that relates the resolvent of the Hermitianization of a data matrix to that of its hidden signal matrix in Section \ref{subsec:subordination}. We next employ this subordination relation to derive expressions for the overlap between data singular vectors and the signal singular vectors in Section \ref{subsec:Singular Vector Overlaps}. We then apply these results to study the extensive spike model in which the rank of the signal spike is assumed to grow linearly with the number of variables (and observations) in Section \ref{subsec:SVD of extensive spike}. There we map out the phase diagram of the SVD as a joint function of signal strength and rank ratio. Intriguingly, we find an that certain transitions in the singular value spectrum of the data do {\it not} coincide with the detectability of the signal, as they do in the finite-rank model. Finally, in Section \ref{sec:optimal rie} we exploit the expressions for singular vector overlaps in order to derive optimal rotationally invariant estimators for both data denoising (Section \ref{subsec:denoising}) and signal-covariance estimation (Section \ref{subsec:cov estimation}). We find that unlike in the finite-rank model, in the extensive-rank model signal singular values interact nontrivially to generate data singular values. Therefore, we obtain more sophisticated optimal data denoisers and signal-covariance estimators that take into account these nontrivial extensive-rank interactions, and which furthermore significantly outperform their simpler, non-interacting, finite-rank counterparts.  

We note that during the preparation of this manuscript, a set of partially overlapping results appeared on a pre-print server \cite{Pourkamali2023}. In our discussion section, we describe the relation and additional contributions of our work relative to that of \cite{Pourkamali2023}.

\section{A Motivation: Inadequacies of the Finite-Rank Spiked Matrix Model}\label{sec:Spiked Matrix Model}

Let $Y$ be an $N_1\times N_2$ signal matrix. We can think of each of the $N_1$ rows of $Y$ as a variable, and each of the $N_2$ columns as a distinct experimental condition or time point, with $Y_{ij}$ representing the clean, uncorrupted value of variable $i$ under condition $j$. Now consider a noisy data matrix $R$, given by
\begin{equation}
  R=Y+X,
  \label{eq:R=Y+X}
\end{equation}
where $X$ is a random $N_1\times N_2$ additive noise matrix. $X$ is assumed to have well-defined limiting singular value spectrum in the limit of large $N_1$ with fixed aspect ratio, $c=\nicefrac{N_1}{N_2}$. Furthermore we assume the probability distribution $P_X(X)$ over $X$ is rotationally invariant, meaning $P_X(X) = P_X(O_1 X O_2)$ where $O_1$ and $O_2$ are orthogonal matrices of size $N_1\times N_1$ and $N_2\times N_2$ respectively. These assumptions guarantee the asymptotic freeness of $X$ and $Y$. For a general definition of freeness, see \cite{Mingo2017}.

We are interested in understanding the relationship between the singular value decomposition (SVD) of the data matrix $R$ and the SVD of the clean signal matrix $Y$. In general we will write the SVD of the data as 

\begin{equation}
  R=\hat{U}_1\hat{S}\hat{U}_2^T=\sum_k \hat{s}_k\hat{\bm{u}}_{\bm{1}k}\hat{\bm{u}}_{\bm{2}k}^T.
\end{equation}
where each $\hat{U}_a$, for $a=1,2$, is an $N_a\times N_a$ matrix with orthonormal columns, $\hat{\bm{u}}_{\bm{a}{k}}$ for $k=1...N_a$, and $\hat{S}$ is a diagonal $N_1\times N_2$ matrix with $\hat{s}_k$ along the diagonal.

As a motivating example, we will study a version of the spiked matrix model \cite{Baik2005,Loubaton2011,Benaych-Georges2012} in which the signal matrix $Y$ is given by 
\begin{equation}
  Y=sU_1U_2^T=s\sum_{k=1}^K \bm{u}_{\bm{1}k}\bm{u}_{\bm{2}k}^T. ,
\end{equation}
where each $U_a$, for $a=1,2$, is an $N_a\times K$ matrix with orthonormal columns, $\bm{u}_{\bm{a}{k}}$ for $k=1...N_a$, and $s$ is the signal strength. This signal model can be thought of as a rank $K$ spike of strength $s$ in that its singular value spectrum has $K$ singular values all equal to $s$. 

\begin{figure}
  \centering
  \includegraphics{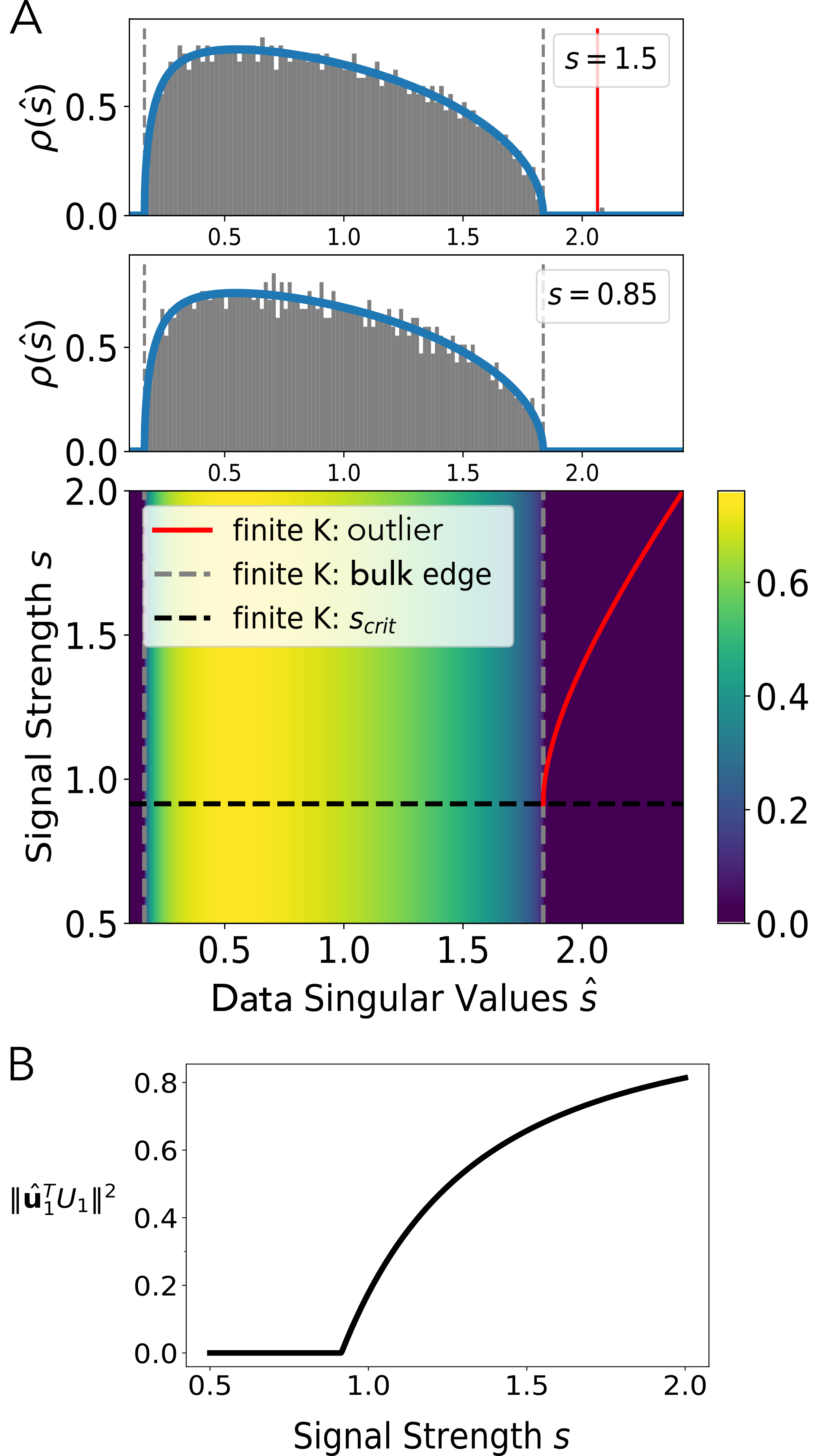}
  \caption{\textbf{Background: Signal-Detectability Phase Transition in the Finite-Rank Spiked Matrix Model. A.} The color plot in the bottom panel shows the singular value spectrum of the spiked matrix model given by the finite-rank theory in the asymptotic limit with aspect ratio, $c=\frac{N_1}{N_2}=0.7$ (see Appendix \ref{sec:Finite-Spike-Appendix} for formulas). The singular value of the data matrix, $R$, is in the x-axis, and the strength of the single non-zero singular value of the signal matrix, $Y$, is in the y-axis. The ``bulk'' spectrum of the data is identical to the spectrum of the noise matrix, $X$. The bounds of that spectrum are the vertical dashed grey lines. Above the critical signal, $s_{crit} = c^{\nicefrac{1}{4}}$ (black horizontal), the data has an outlier singular value shown as a solid red curve. The top two panels show theory curves corresponding to two horizontal slices, with $s=0.85, 1.5$, together with a histogram of singular values each of a single instantiation with $N_2=2000$. The top panel has a single outlier eigenvalue very close to the theory prediction. The panel below shows a data spectrum that is indistinguishable from noise. \textbf{B.} The overlap of the top left singular vector of the data with the left singular subspace of the signal, given by the finite-rank theory. The overlap becomes non-zero at exactly the same critical signal, $s_{crit}$, at which an outlier singular value appears in the data. $X$ is Gaussian i.i.d. with variance $\nicefrac{1}{N_2}$ throughout. }
  \label{fig:Fig1_background_spike_model}
\end{figure}

In the finite-rank setting, where $K$ remains finite as $N_1,N_2 \rightarrow \infty$, there is a signal-detectability phase transition \cite{Baik2005,Benaych-Georges2012} in the singular value structure of the data matrix $R$. For $s<s_{crit}$, where $s_{crit}$ is a critical signal strength that depends on the singular value spectrum of the noise matrix $X$, the entire signal in $Y$ is swamped by the additive noise $X$ and cannot be seen in the data $R$. More precisely, in the large size limit, when $s<s_{crit}$ the singular value spectrum of the data $R$ is {\it identical} to the singular value spectrum of the noise $X$. Furthermore, {\it no} left (right) singular vector of the data matrix $R$ has an $O(1)$ overlap with the $K$-dimensional signal subspace corresponding to the column space of $U_1$ ($U_2$). However, for $s>s_{crit}$ the singular value spectrum of the data $R$ is now not only composed of a noise bulk, identical to the spectrum of $X$, as before, but also acquires $K$ outlier singular values all equal to $\hat s$. The location of the data spike at $\hat s$, occurs at a slightly larger value than the signal spike at $s$. This reflects singular-value inflation in the data $R$ relative to the signal $Y$, due to the addition of noise $X$. Furthermore, each singular vector of the data $R$ corresponding to an outlier singular value acquires a nontrivial $O(1)$ overlap with the $K$ dimensional signal subspace of $Y$ even in the asymptotic limit $N_1,N_2 \rightarrow \infty$. 

The location of the outlier data singular values and their corresponding singular-vector overlaps with the signal subspace have been calculated for {\it finite} $K$ and general rotationally invariant noise matrices $X$ \cite{Benaych-Georges2012}. In the special case where the elements of $X$ are i.i.d. Gaussian, explicit formulas can be derived (see Appendix \ref{sec:Finite-Spike-Appendix} for a review). This signal-detectability phase transition in the finite-rank spiked model is depicted in Fig. \ref{fig:Fig1_background_spike_model} for an i.i.d. Gaussian noise matrix $X$.

Notably, according to the finite-rank theory, the $K$ spikes do not interact. More precisely, above the critical signal strength, in the large-size limit, the $K$ identical singular values of $Y$ are all predicted to map to $K$ identical outlier singular values of the data matrix $R$. Furthermore, the overlaps of the $K$ corresponding data singular vectors with the signal subspace are predicted to be identical and completely independent of the finite value of $K$ (see \cite{Bai2008_finiteK_fluctuations} however, for finite-size fluctuations in the square-symmetric spiked covariance model). More generally, if the signal $Y$ consists of $K$ {\it different} rank $1$ spikes each with a unique signal strength $s_l$ for $l=1,\dots,K$, the corresponding location of the data spike $\hat s_l$ can be computed by inserting each $s_l$ into a single local singular value inflation function $\hat s(s)$ (depicted in Fig. \ref{fig:Fig1_background_spike_model}), without considering the location of any other signal spike $s_{l'}$ for $l' \neq l$. In this precise sense, at finite $K$ the spikes do not interact; one need not consider the position of any other signal spikes to compute how any one signal spike is inflated to a data spike. The same non-interacting picture is true for singular vector overlaps (Fig. \ref{fig:Fig1_background_spike_model}B).

This lack of interaction between different spikes in the signal as they are corrupted to generate data spikes, allows optimal denoising operations based on the finite-rank theory to be remarkably simple.  For example, estimators for both data denoising \cite{Shabalin2013,Gavish2014, Gavish2017}, which corresponds to trying to directly estimate the signal $Y$ given the corrupted data $R$, and covariance estimation \cite{Donoho2018}, which corresponds to estimating the true covariance matrix $C = YY^T$ from the data $R$, both involve applying a {\it single} shrinkage function, that non-linearly modifies each data singular value of $R$ in a manner that acts \textit{independently} of any other singular value. This shrinkage function, applied to each data singular value $\hat s$, in a sense optimally undoes the singular-value inflation $s \rightarrow \hat{s}$ and compensates for the singular-vector rotation $\bm{u_a}\rightarrow \hat{\bm{u}}_{\bm{a}}$ which arise in going from signal $Y$ to data $R=Y+X$. Moreover, the reason the shrinkage can act independently on each data singular value is directly related to the property of the finite-rank theory that each signal singular value is inflated {\it independently} through the same inflation function, while each signal singular vector is rotated {\it independently} through the same random rotation.

\begin{figure*}
  \centering
  \includegraphics[width=\textwidth]{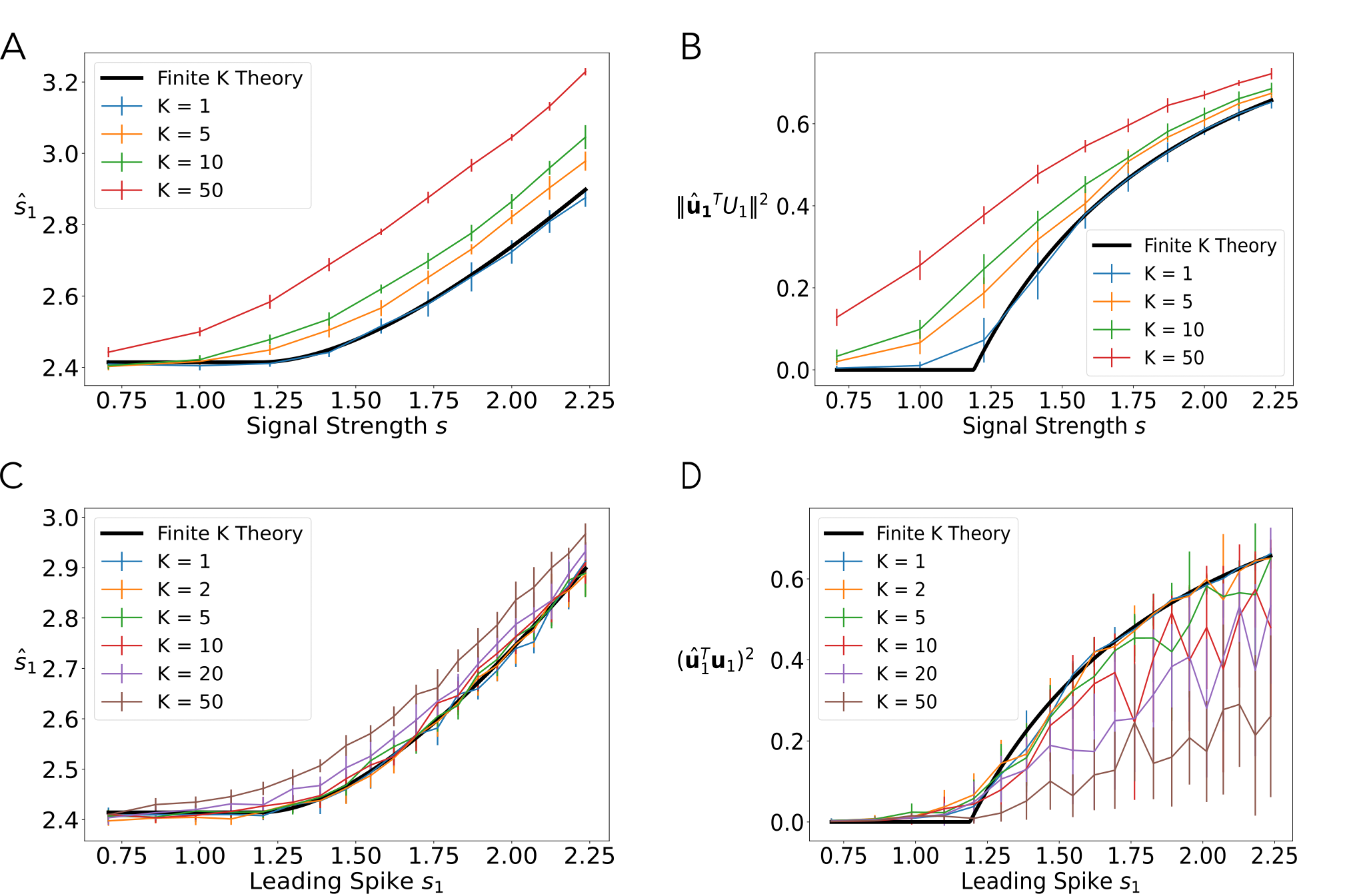} 
  \caption{\textbf{Finite-Rank Theory Fails To Capture the Singular Value Structure of the Spiked Rectangular Matrix Model. A.} Singular-value inflation, i.e. leading data singular value, $\hat s_1$, as a function of signal singular value, $s_1$, for spikes of various ranks $K$. Black shows the finite-rank theory (which is independent of the rank of a spike). Matrix size in this and all subsequent panels is $N_1=1000$, $N_2=500$. Numerical results are presented as mean and standard deviation over $10$ instantiations for each value of $b$ and $s$. \textbf{B.} Singular-vector rotation, i.e. overlap of leading data left singular vector, $\hat{\mathbf{u}}_1$ with the $K$-dimensional left singular space of the signal, $U_1$. \textbf{C.} To break the degeneracy of the spikes with rank $K>1$ in panel A, here we consider a single leading signal spike with singular value $s$, along with $K-1$ spikes drawn independently and uniformly in $\lrb{0,s}$. We plot the leading data singular value as a function of $s$ for various $K$, compared to finite-rank theory (black) \textbf{D.} For the same signal model as in panel (C), we plot the overlap of the leading data left singular vector with the leading signal singular vector as function of signal strength $s$, for different $K$ and for the finite-rank theory (black). We see that the finite-rank theory incorrectly estimates both the singular values and singular vectors of signals of even moderate rank $K$. See Appendix \ref{sec:Finite-Spike-Appendix} for finite-rank theory formulas.} 
  \label{fig:Figure 2 finite-rank theory fails}
\end{figure*}

In this work, however, we find that the assumptions and resulting consequences of the finite-rank theory may constitute a significant limitation for the practical application of this model to both explain the properties of noise corrupted data, as well as to optimally denoise such data. To illustrate, we test the finite-rank theory for various values of $K$, with $N_1$ and $N_2$ fixed. In Figure \ref{fig:Figure 2 finite-rank theory fails} we show simulation results in which we find substantial deviations between simulations and finite-rank theory predictions, for both the location of the leading data singular value outlier, and the data-signal singular-vector overlap, for $K$ as small as $10$ with $N_1=1000$. Thus, even for moderate numbers of spikes and relatively large matrices, the finite-rank theory cannot explain the SVD of the data well, (though as mentioned above, see \cite{Bai2008_finiteK_fluctuations} for finite-size fluctuations in the square-symmetric case). As a consequence, as we will show below, typical denoising techniques, which depend crucially on the predicted singular structure of the data, perform poorly, even for moderate $K$.

Thus, motivated by the search for better denoisers of higher-rank data, we extend the finite-rank theory to a completely different asymptotic limit of extensive rank, in which the rank $K$ of the data is proportional to the number of variables $N_1$ as both become large.  We show that our extensive-rank theory both: (1) more accurately explains the SVD of large data matrices of even moderate rank, and (2) provides better denoisers in these cases, than the finite-rank theory. And interestingly, our extensive-rank theory reveals qualitatively new phenomena that do not occur at finite-rank, including highly nontrivial interactions between the extensive number of signal singular values, as they become corrupted to generate data singular values, under additive noise. 

\section{Mathematical Preliminaries}\label{sec:mathematical prelims}

We review some basic concepts from random matrix theory and introduce notation. Let $M$ be an $N\times N$ Hermitian matrix $M$. We denote by $G_M(z)$ the matrix resolvent of $M$:
\begin{equation}
  G_M(z) := (zI-M)^{-1}.
\end{equation}
We define the normalized trace operator $\tau$ as 
\begin{equation}
  \tau\lrb{M}:=\frac{1}{N}\trace{}\lrb{M}.
\end{equation}
The Stieltjes transform $g_M(z)$ is the normalized trace of $G_M(z)$:
\begin{equation}
  g_M(z) := \tau \left[(zI-M)^{-1}\right].
\end{equation}

In this work, we will be interested in the singular values and vectors of rectangular matrices. In order to apply Hermitian matrix methods to a rectangular matrix $R\in\mathbb{R}^{N_1\times N_2}$, we will work with its Hermitianization, which we denote with boldface throughout:
\begin{equation}\label{eq-define-Hermitianization}
\bm{R}:=\left[\begin{array}{cc}
0 & R\\
R^T & 0
\end{array}\right],
\end{equation}
which is an $N\times N$ Hermitian matrix, with $N = N_1+N_2$. The eigenvalues and eigenvectors of $\bm{R}$ can be written $\pm s,\frac{1}{\sqrt{2}}\left(\begin{array}{c}
\bm{u_1}\\
\pm \bm{u_2}
\end{array}\right)$, where $s$ is a singular value of $R$, and $\bm{u_1},\bm{u_2}$ are the corresponding left and right singular vectors. This will allow us to extract information about the singular value decomposition of a rectangular matrix $R$ from the eigen-decomposition of the Hermitian matrix $\bm{R}$. 

Hermitianization leads naturally to a Hermitian {\it block resolvent}, which is a function of two complex scalars $z_1$ and $z_2$, rather than one: 
\begin{equation}
\bm{G}^{{R}}\left(\bm{z}\right):=\mat2{z_1 I_{N_1}}{-R}{-R^T}{z_2 I_{N_2}}^{-1},
\end{equation}
where $\bm{z}=\lr{z_1,z_2}$ is a complex vector. This block resolvent can be computed explicitly, with each block written in terms of a standard square-matrix resolvent:
\begin{equation}\label{eq-define-block-resolvent}
\bm{G}^{{R}}\left(\bm{z}\right)=\left[\begin{array}{cc}
z_2 G_{RR^{T}}\left(z_1 z_2\right) & RG_{R^{T}R}\left(z_1 z_2\right)\\
R^{T}G_{RR^{T}}\left(z_1 z_2\right) & z_1 G_{R^{T}R}\left(z_1 z_2\right)
\end{array}\right].
\end{equation}

Analogously, we define the {\it block} Stieltjes transform $\bm{g}^R(\bm{z})$ as the $2$-element complex vector consisting of the normalized traces of each diagonal block of $\bm{G}^{{R}}$: 
\begin{subequations}
  \begin{align} \label{eq-block-Stieltjes}
  g_{1}^{R}\left(\bm{z}\right)&=\tau_1\left[{G}_{11}^{R}\left(\bm{z}\right)\right] =z_{2}g_{RR^{T}}\left(z_{1}z_{2}\right)\\
  g_{2}^{R}\left(\bm{z}\right)&= \tau_2\left[{G}_{22}^{R}\left(\bm{z}\right)\right]=z_{1}g_{R^{T}R}\left(z_{1}z_{2}\right).
\end{align}

\end{subequations}
Here we have introduced notation for the block-wise normalized traces: 
\begin{equation}
  \tau_a\lr{M}:=\frac{1}{N_a}\trace{}\lrb{M_{aa}},
\end{equation}
where $M_{aa}$ is the $a$th diagonal block of size $N_a\times N_a$.

Notationally, we write the block vectors and block matrices $\bm{g}^R$ and $\bm{G}^R$ in bold, while we indicate the component blocks in standard roman font, with the indices $a,b$ for both scalar, $g_a^R$, and matrix $G_{ab}^R$ blocks, with $a,b\in\left\{1,2\right\}$. 

We will also use the fact that the eigenvalues of $RR^T$ and $R^TR$ differ by exactly $\left|N_1-N_2\right|$ zeros, implying the two elements of $\bm{g}^R$ are related by $g_2^R\lr{\bm z} = \frac{z_1}{z_2}cg_1^R\lr{\bm z} + \frac{1-c}{z_2}$.

Each element $g_a^R\lr{z}$ can be written in terms of the corresponding singular value density:
\begin{subequations}
  
\begin{align} 
  g_{1}^{R}\left(z_{1},z_{2}\right)&=\int_{-\infty}^{+\infty}\frac{z_2}{z_1z_2 - s^2}\rho_1^R\lr{s}\rm{d}s\\
  g_{2}^{R}\left(z_{1},z_{2}\right)&=\int_{-\infty}^{+\infty}\frac{z_1}{z_1z_2 - s^2}\rho_2^R\lr{s}\rm{d}s,
\end{align}
\end{subequations}
where $\rho_a^R(s)$ denotes the singular value distribution of $R$, accounting for $N_a$ singular values. Note that for non-zero $s$ with finite singular value density, $\rho_2^R\lr{s}=c\rho_1^R\lr{s}$.

The special case in which the two arguments are equal, $z_1=z_2=z$, will be important and so we abbreviate: $\bm{g}^R(z) :=\bm{g}^R(z,z)$.

We can write an inversion relation for the singular value densities using the Sokhotski-Plemelj theorem, which states, $\xlim{\eta}{0^+}\im\lrb{\int\frac{f\lr{x}}{x-i \eta}\rm dx}=\pi f\lr{0}$. Applying this theorem to $f\lr{x}=\frac{z}{z+x}\rho_a^R\lr{x}$ yields:
\begin{equation}\label{eq-inversion-relation}
\rho_{a}^{R}\left(s\right)=\frac{2}{\pi}\xlim{\eta}{0}\mathrm{Im}\lrb{g_{a}^{R}\left(s-i\eta\right)}.
\end{equation}

Finally, we define the {\it block} $\xr$-transform: 
\begin{equation}\label{eq-R-transform}
\bm{\xr}^R(\bm{t})=\left(\bm{g}^R\right)^{-1}(\bm{t})-\frac{1}{\bm{t}},
\end{equation}
where $\bm{t}\in\mathbb{C}^2$ is in the range of $\bm{g}^R$;  we denote by $\left(\bm{g}^R\right)^{-1}$ the functional inverse of the block Stieltjes transform $\bm{g}^R$, satisfying $\left(\bm{g}^R\right)^{-1}\lr{\bm{g}^R\lr{\bm z}}=\bm z$; and $\nicefrac{1}{\bm{t}}$ is the component-wise multiplicative inverse of $\bm{t}$. 

The block $\xr$-transform will arise naturally in our calculation of the subordination relation for the sum of free rectangular matrices, $R=Y+X$, and as we shall verify, it is additive for independent, rotationally invariant matrices:
\begin{equation}
  \bm{\xr}^R\lr{\bm t} = \bm{\xr}^Y\lr{\bm t} + \bm{\xr}^X\lr{\bm t}.
\end{equation}

\section{The Singular Value Decomposition of Sums of Rectangular Matrices}\label{sec:SVD of Sums}

In this section, we characterize how an additive noise matrix $X$ deforms the singular values and vectors of a signal matrix $Y$ to generate singular values and vectors of the data matrix $R=Y+X$ (see \eqref{eq:R=Y+X} and following text). We consider general signal matrices of the form 
\begin{equation}
  Y=U_1SU_2^T,
\end{equation}
where each $U_a$ is an $N_a\times N _a$ orthonormal matrix ($a=1,2$), and $S$ is $N_1\times N_2$ diagonal matrix. 

We begin by deriving an asymptotically exact subordination formula relating the block resolvents \eqref{eq-define-block-resolvent} of $R$ and $Y$ in the limit $N_1,N_2\to\infty$ with the aspect ratio $c=\nicefrac{N_1}{N_2}$ fixed. From this, we extract both the singular value spectrum of $R$, as well as the overlaps between the singular vectors of $R$ and those of the signal matrix, $Y$.

\subsection{A Subordination Relation for the Sum of Rectangular Matrices}\label{subsec:subordination}

Exploiting the rotational invariance of $P_X(X)$, we first calculate the block resolvent of $R$ as an expectation over arbitrary rotations of the noise $X$. Thus, we write $R=Y+O_1XO_2^T$, where $O_a$ are Haar-distributed orthogonal $N_a\times N_a$ matrices. We can write the Hermitianization \eqref{eq-define-Hermitianization} of $\bm R$ in terms of the Hermitianized $\bm X$ and $\bm Y$:
\begin{equation}
  \bm{R} = \bm{Y} + \bar{\bm{O}}\bm{X}\bar{\bm{O}}^T,
\end{equation}
where we have written $\bar{\bm{O}}=\mat2{O_1}{0}{0}{O_2}$.

The main result of this section is the following subordination relation for the expectation of the block resolvent $\bm{G}^{R}$, taken over the random block-orthogonal matrix $\bar{\bm{O}}$.
\begin{equation}
\label{eq-subordination-main}
  \eO\lrb{\bm{G}^{R}(z)} = \bm{G}^{Y}\left(z-\bm{\xr}^{X}\left(\bm{g}^{R}\left(z\right)\right)\right).
\end{equation}
As mentioned above, this notation refers to the special case in which the argument to $\bm{g}^R$ is the two-dimensional complex vector with equal arguments, $z_1=z_2=z$. Note that the argument to $\bm{G}^Y$, by a slight abuse of notation, is the vector $z-\xr{R}_a^X\lr{\bm{g}^R\lr{z}}$ for $a=1,2$. In Appendix \ref{sec:Annealed-Appendix} we present the detailed derivation for this case, which is sufficient for computing the singular values and associated singular-vector overlaps. We provide a sketch of the calculation here. The general case follows.

We first write the analog of a partition function,
\begin{equation}
  \cZ^R\lr{\bm Y}=\det\lr{zI-\bm R}^{-
  \nicefrac{1}{2}},
\end{equation}
and observe that we can write the desired matrix inverse as a derivative of the corresponding free-energy:
\begin{equation}
  \bm{G}^R\lr{z} = 2\frac{\rm d}{\rm d \bm{Y}}\log\cZ^R\lr{\bm Y}.
\end{equation}

We would like to average this over the block-orthogonal matrix $\bar{\bm O}$, yielding a ``quenched'' average free-energy. In Appendix \ref{sec:Justify-Annealed-Appendix}, we show that in the large $N$ limit, the quenched and annealed averages are equivalent. In short, viewing $\log\cZ^R$ as a function of $\bar{\bm O}$, we find it has Lipschitz constant proportional to $\nicefrac{1}{\sqrt N}$, and then use the concentration of measure of the orthogonal group, $\S\O\left(N\right)$, with additional concentration inequalities to show that:
\begin{multline}
  \xlim{N}{\infty}\frac{1}{N}\eO\lrb{\log\cZ^R\lr{\bm Y}} = \\ \xlim{N}{\infty} \frac{1}{N}\log\eO\lrb{\cZ^R\lr{\bm Y}}.
\end{multline}
We can therefore calculate our desired block resolvent as
\begin{equation}
    \eO\lrb{\bm{G}^{R}(z)}=2\frac{\rm d}{\rm d \bm{Y}}\log\eO\lrb{\cZ^R\lr{\bm Y}}.
\end{equation}
We proceed by writing the determinant as a Gaussian integral, 
\begin{equation}\label{eq-main-gaussian-integral}
  \cZ^R\lr{\bm Y} = \int \frac{d\mb{v}}{\lr{2\pi}^{\nicefrac{N}{2}}} \exp\lr{-\frac{1}{2}\mb{v}^T\left(z I-\bm R\right)\mb{v}},
\end{equation}
and then we substitute $\bm R=\bm Y +\bar{\bm O}\bm X \bar{\bm O}^T$, extract terms that do not depend on $\bar{\bm O}$, and take the expectation of the terms that do, which yields an intermediate integral,
\begin{equation}\label{eq-HCIZ-main}  I^X\lr{\mb{v}}\equiv\eO\lrb{\ex{\frac{1}{2} \mb{v}^T\bar{\bm O}\bm X\bar{\bm O}^T\mb{v}}},
\end{equation}

This integral is analogous to the Harish-Chandra-Itzykson-Zuber (HCIZ) or spherical integral, which appears in the calculation of the subordination relation for sums of square-symmetric matrices \cite{Bun2016,Bun2017,potters_bouchaud_2020}. 
We compute this ``block-spherical'' integral asymptotically in Appendix \ref{sec:rank1-HCIZ-appendix}, and highlight key points of the calculation here. 

First, we observe the key difference between our calculation and the square-symmetric case. In the square symmetric case, the expectation is over a single Haar-distributed orthogonal matrix that rotates $\bm{v}$ arbitrarily, and so the expectation depends only on the norm of $\bm{v}$. In our rectangular case, however, $\bar{\bm{O}}$ has two blocks, and they rotate the $N_1$- and $N_2$-dimensional blocks of $\bm{v}$ separately, so that $I^X\lr{\bm{v}}$ depends on the norms of each of these two blocks. Therefore, we define the $2$-component vector, $\bm t$ with components
\begin{equation}
  t_a:=\frac{1}{N_a}\norm{\bm{v}_a}^2.
\end{equation}
We calculate the expectation \eqref{eq-HCIZ-main} by performing an integral over an arbitrary $N$-dimensional vector, while enforcing block-wise norm constraints using the Fourier representation of the delta function, and introducing integration variables, $q_1$ and $q_2$.

To compute the integral, we make a saddle-point approximation in the asymptotic limit of large $N$. Appealingly, we find the saddle-point conditions are of the form:
\begin{subequations}
\begin{align}
  t_1 &= q_2^*g_{XX^T}\lr{q_1^* q_2^*} \\
  t_2 &= q_1^*g_{X^TX}\lr{q_1^* q_2^*}.
\end{align}
\end{subequations}
That is, the block Stieltjes transform, $\bm{g}^X\lr{\bm{q}^*}=\bm{t}$, arises naturally, and the saddle-point of the block-spherical integral \eqref{eq-HCIZ-main} is its functional inverse evaluated at the vector of block-wise norms of $\bm v$.

Inserting the saddle-point solution, we find that asymptotically
\begin{equation}
  I^X\lr{\bm{v}} = \exp\lr{\frac{{N}}{2}H^X\lr{\bm t}},
\end{equation}
where, for a neighborhood of values of $\bm{t}$ around $0$, the saddle-point free energy itself, $H^X\lr{\bm{t}}$, has gradient with elements proportional to the block $\xr$-transform \eqref{eq-R-transform}:
\begin{equation}
  \frac{\rm d H^X\lr{\bm t}}{\rm{d} t_a} = \frac{N_a}{N}\xr_a^X\lr{\bm t}.
\end{equation}
Thus, the block $\xr$-transform arises via the anti-derivative of the logarithm of the block-spherical integral, analogously to the regular $\xr$-transform in the case of square-symmetric matrices.

Note that given the definition in \ref{eq-HCIZ-main}, it is straightforward to see that $I^R\lr{\bm v}=I^Y\lr{\bm v} I^X\lr{\bm v}$, and thus $H^R\lr{\bm t}=H^Y\lr{\bm t} + H^X\lr{\bm t}$. Therefore, we have established the additivity of the block $\xr$-transform as well.

Continuing with the derivation of the subordination relation \eqref{eq-subordination-main}, we next substitute the result for $I^X\lr{\bm v}$ back into the Gaussian integral over $\bm v$ \eqref{eq-main-gaussian-integral}, and then introduce another pair of integration variables, $\hat{\bm t}$, in order to decouple $\bm v$ from its block-wise norms, $\bm t$. Performing the Gaussian integral we find
\begin{equation}
    \e_{\bar{\bm{O}}} \lrb{\cZ^{R}\lr{\bm{Y}}} \propto \int \rm{d}\bm{t} \rm{d}\hat{\bm{t}}\exp\lr{\frac{N}{2}P^{X,Y}\lr{\bm t,\hat{\bm t}} }, 
\end{equation}
with
\begin{align}\label{eq-final-potential-main}
  P^{X,Y}\lr{\bm t,\hat{\bm t}} := &-\frac{1}{N}\lr{N_1 t_1 \hat{t}_1 +N_2 t_2\hat{t}_2} + H^X\lr{\bm t} \nonumber \\ 
  &-\frac{1}{N}\log\det\bm{G}^Y\lr{z-\hat{\bm t}}. 
\end{align}

Note that the block resolvent of $Y$ arises here naturally as a function of the two-element vector, $z-\hat{\bm t}$, despite the fact that we set out to find $\bm G^R$ evaluated at the point $\lr{z,z}$.

The integrals over $\bm t$ and $\hat{\bm t}$ yield an additional pair of saddle-point conditions. The first requires $\hat{\bm t}^*=\bm{\xr}^X\lr{\bm t^*}$ and combining with the second gives
\begin{equation}\label{eq-final-saddle-point}
  \bm{t}^* = \bm{g}^Y\lr{z - \bm{\xr}^X\lr{\bm t^*}}.
\end{equation}
We have thus found the desired annealed free energy, $2\log\eO\lrb{\cZ^R}=NP^{X,Y}\lr{\bm t^*,\bm{\xr}^X\lr{\bm t^*}}$ (see \eqref{eq-final-potential-main}).

We next take the derivative with respect to $\bm Y$ (see Appendix \ref{sec:Annealed-Appendix} for a more careful treatment), which gives 
\begin{equation}
\eO\lrb{\bm{G}^R\lr{z}} = \bm{G}^Y\lr{z-\bm{\xr}^X\lr{\bm t^*}}.  
\end{equation}
Finally to find $\bm t^*$, we take the block-wise normalized traces to find $\bm g^R\lr{z} = \bm g^R\lr{z-\bm R^X\lr{\bm t^*}}=\bm t^*$, and that completes the derivation of the block resolvent subordination relation \eqref{eq-subordination-main}.

We note that the saddle-point condition \eqref{eq-final-saddle-point} turns out to be the subordination relation for the block Stieltjes transform:
\begin{equation}
\label{eq-subordination-stieltjes}
\bm{g}^R\lr{z} = \bm{g}^Y\lr{z - \bm{\xr}^X\lr{\bm{g}^R\lr{z}}}. 
\end{equation}

Note that while $\bm{g}^R$ is evaluated at the scalar point $(z,z)$, the argument to $\bm{g}^Y$ is the vector subordination function $\bb{\zeta}\equiv z-\bm{\xr}^{X}\left(\mathbf{g}^{R}\left(z\right)\right)$ whose two components are distinct in general.

The singular value spectrum of the sum of rectangular matrices can thus be obtained by first finding the block Stieltjes transform, either by employing the additivity of the block $\xr$-transform or by solving the subordination relation \eqref{eq-subordination-stieltjes}, and then using the inversion relation \eqref{eq-inversion-relation}.

\subsection{Deformation of Singular Vectors \\ Due to Additive Noise}\label{subsec:Singular Vector Overlaps}

Turning now to the singular vectors of the data matrix $R=\hat{U}_1\hat{S}\hat{U_2}^T=Y+X$, we quantify the effect of the noise, $X$, on the signal, $Y=U_1SU_2^T$, via the matrix of squared overlaps between the clean singular vectors of the signal, $U_a$ with $a=1,2$ for left and right, respectively, and the noise-corrupted singular vectors of the data, $\hat{U}_a$, written as $\lr{\hat{U}_a^TU_a}^2$.

In the noiseless case $X=0$, one has $\lr{\hat{U}_a^TU_a}^2 = I_{N_a}$, signifying perfect correspondence between signal and data singular vectors. In the presence of substantial noise, the overlaps of a signal singular vector are generically distributed over order $N_a$ data singular vectors and are of order $\nicefrac{1}{N_a}$, therefore we define the rescaled expected square overlap between a given singular vector, $\hat{\bm{u}}_{\bm a}$ of $R$ with corresponding singular value $\hat s$, and a given singular vector, $\bm{u}_{\bm a}$ of $Y$, with corresponding singular value $s$, where once again $a=1,2$ for left and right singular vectors, respectively:
\begin{equation}
  \Phi_a\lr{\hat s, s} = N_a\e\lrb{\lr{\hat{\bm{u}}_{\bm a}^T\bm{u}_{\bm a} }^2}.
\end{equation}

To see how to obtain the expected square overlaps from the block resolvent, $\bm{G}^R\lr{z}$, we write each of the diagonal blocks, $G_{aa}\lr{z}^R$ \eqref{eq-define-block-resolvent}, in terms of their eigen-decomposition, and multiply on both sides by a ``target'' singular vector of $Y$, say $\bm{u}_{\bm{a}}$ with associated singular value $s$:

\begin{equation}
  \bm{u}_{\bm{a}}^TG^R_{aa}\lr{z}\bm{u}_{\bm{a}} = \sum_{k=1}^{N_a}\frac{z}{z^2-\hat s_k^2}\lr{\bm{u}_{\bm{a}}^T\hat{\bm{u}}_{\bm{a}k}}^2.
\end{equation}
If we choose $z=\hat s-i\eta$ where $\rho_a^R\lr{\hat s}\sim O\lr{1}$, with $N_a^{-1}\ll \eta\ll 1 $, and take the imaginary part, then we get a weighted average of the square overlaps of a macroscopic number of singular vectors of $R$, $\hat{\bm u}_{\bm{a}k}$, that have singular values close to $\hat s$, with the target singular vector $\bm{u}_{\bm a}$, each weighted by $\nicefrac{\pi}{2}\rho_a^R\lr{\hat s_k}$. If we first take the limit of large $N_a$ and then take $\eta\rightarrow 0$ we obtain the expectation:
\begin{equation}
\xlim{\eta}{0}\frac{2}{\pi}\im\lrb{\bm{u}_{\bm{a}}^TG^R_{aa}\lr{\hat s- i\eta}\bm{u}_{\bm{a}}}\rightarrow \rho_a^R\lr{\hat s}\Phi_a\lr{\hat s,s}.
\end{equation}

Now, we use the subordination relation \eqref{eq-subordination-main} to replace the resolvent of $R$ with the resolvent of $Y$: $G_{aa}^R\lr{z}=G_{aa}^Y\lr{\bm \zeta\lr{z}}$ where we have written the $2$-component vector
\begin{equation}
  \bm{\zeta}(z)=z -\bm{\xr}^{X}\left(\mb{g}^{R}\left(z\right)\right).
\end{equation}
Since $\bm{u}_{\bm{a}}$ is an eigenvector of ${G}^Y_{aa}\lr{\zeta_1,\zeta_2}$ with eigenvalue $\frac{\zeta_b }{\zeta_1\zeta_2-s^2}$ where $b=2$ for $a=1$ and $b=1$ for $a=2$, we find
\begin{subequations} \label{eq-overlap-both-formulas}
\begin{align}
\Phi_1\lr{\hat s, s}&= \frac{2}{\pi \rho^R_1\lr{\hat s}} \lim_{\eta\to0}\im\frac{\zeta_{2}\left(\hat{s}-i\eta\right)}{\zeta_{1}\left(\hat{s}-i\eta\right)\zeta_{2}\left(\hat{s}-i\eta\right)-i\eta-s^{2}} \label{eq-overlap-formula-1}\\ 
\Phi_2\lr{\hat s, s} &= \frac{2}{\pi \rho^R_2\lr{\hat s}} \lim_{\eta\to0}\im\frac{\zeta_{1}\left(\hat{s}-i\eta\right)}{\zeta_{1}\left(\hat{s}-i\eta\right)\zeta_{2}\left(\hat{s}-i\eta\right)-i\eta-s^{2}}.\label{eq-overlap-formula-2}
\end{align}
\end{subequations}

These expressions can be written in terms of the real and imaginary parts of the block $\xr$-transform of the noise $X$. In the following section we provide simplified expressions for the important case of Gaussian noise.

\subsubsection{Arbitrary Signal with Gaussian Noise}\label{subsubsec:arbitrary signal gaussian noise}

\begin{figure}
  \centering
  \includegraphics[width=8.6cm, height =14cm]{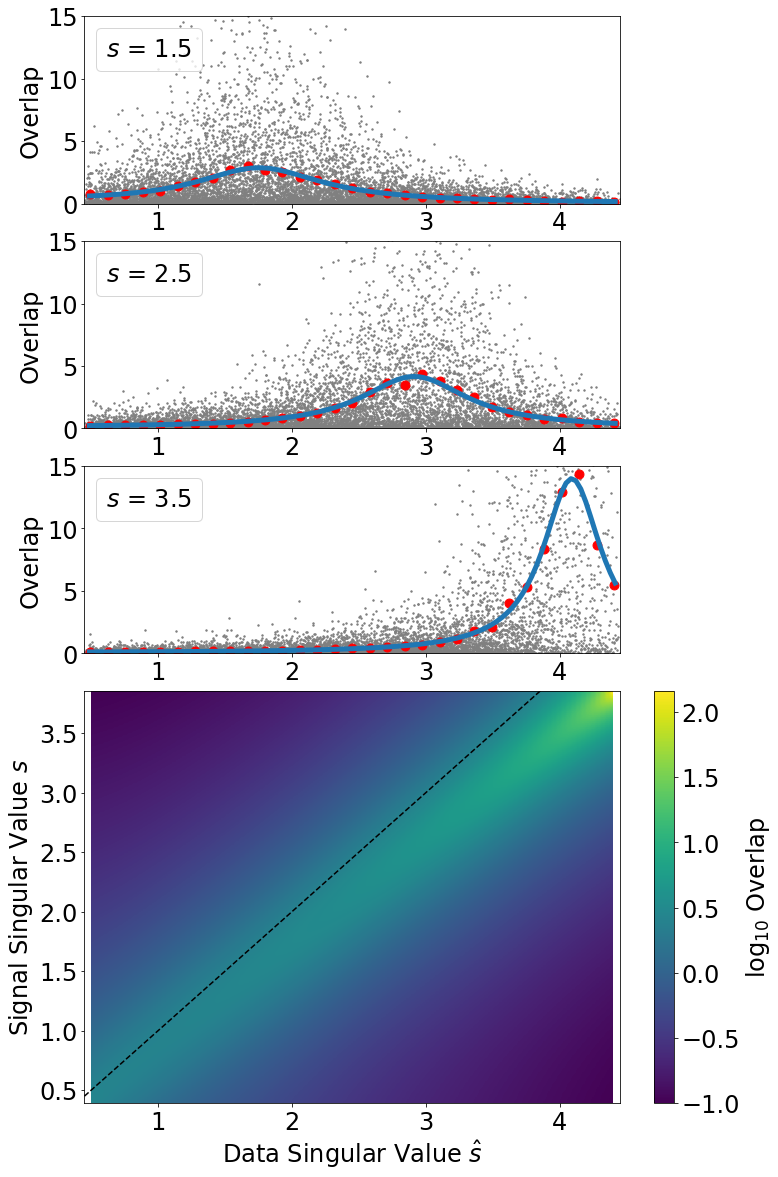}
  \caption{\textbf{Singular Vector Overlaps of Sums of Gaussians.} Color plot (bottom) shows the theoretical prediction of the logarithm of the overlap, $\log_{10} \Phi_1\lr{\hat s, s}$ (Eq \eqref{eq-left-overlap-with-gaussian-noise}), between the left singular vectors of $R=Y+X$ and those of $Y$, as a bivariate function of the associated singular values $\hat s$ of $R$ and $s$ of $Y$. Dashed line is identity, $s=\hat s$. Here the signal $Y$ and noise $X$ are both rectangular Gaussian matrices with aspect ratio $c=\nicefrac{N_1}{N_2}=\nicefrac{3}{2}$. Elements of $Y$ are i.i.d. with variance $\nicefrac{\sigma^2_y}{N_2}$ where $\sigma_y^2=3$. Elements of $X$ are i.i.d. with variance $\nicefrac{\sigma^2_x}{N_2}$ with $\sigma_x^2=1$. Top three panels show singular vector overlaps for $3$ horizontal slices associated with $3$ fixed ``target'' signal singular values $s=1.5,2.5,3.5$, for $10$ realizations of random matrices with $N_1=1500$ and $N_2=1000$. Each grey dot denotes an overlap between a left singular vector of $R$ with singular value $\hat s$ (position on x-axis) with the left singular vector of $Y$ with singular value {\it closest} to $s$. Red dots reflect binning the singular values of $R$ from all $10$ realizations, with number of bins set to $\sqrt N_2$ giving bin width $\approx 0.13$. Blue is the theoretical prediction from $\Phi_1\lr{\hat s, s}$ in \eqref{eq-left-overlap-with-gaussian-noise}. Note that as the signal singular value $s$ increases, $\Phi_1\lr{\hat s, s}$ as a function of $\hat s$ becomes more concentrated about a value {\it larger} than $s$. This reflects the fact that singular vector structure in the signal $Y$ at singular value $s$ is mapped to singular vector structure in the data $R$ at larger singular values $\hat s$, due to singular value inflation under the addition of noise $X$.}
  \label{fig:Figure 3 - Sums of Gaussians}
\end{figure}

We show in Appendix \ref{sec:Gaussian-noise-Appendix} that the block $\xr$-transform of an $N_1\times N_2$ (with $c=\frac{N_1}{N_2}$) Gaussian matrix with i.i.d. entries of variance $\nicefrac{\sigma^2}{N_2}$ is:
\begin{equation}\label{eq-Gaussian-R-transform}
  \bm\xr^X\lr{\bm t} = \sigma^2\cvec2{t_2}{c t_1}.
\end{equation}
Note that from the definition of the $\xr$-transform, one can find that $\xr^A_2\lr{\bm t} t_2=c\xr^A_1\lr{\bm t} t_1$, for any rectangular $A$ with aspect ratio $c$, and \eqref{eq-Gaussian-R-transform} is the only pair of linear functions of $\bm{t}$ that satisfies this constraint.

We substitute \eqref{eq-Gaussian-R-transform} into the block Stieltjes transform subordination relation yielding $\bm g^R\lr{z}=\bm g^Y\lr{\bm\zeta}$ with $\zeta_1 = z - \sigma^2 g_2^R\lr{z}$ and $\zeta_2=z-c\sigma^2g_1^R\lr{z}$, and then use the identity $g^R_2(z)=c g^R_1(z)+\frac{1-c}{z}$ (for arbitrary rectangular $R$, the spectra of $RR^T$ and $R^TR$ differ only by a set of $0$ eigenvalues). Then using the definition $g_1^Y\lr{\bm \zeta}=\zeta_2g_{YY^T}\lr{\zeta_1\zeta_2}$, we arrive at:
\begin{equation}
\label{eq:Arbitrary_Signal_Gaussian_noise}
  g_1^R\lr{z} = \zeta_2\lr{z} g_{YY^T}\left(\zeta_2\lr{z}\left(\zeta_2\lr{z}-\sigma^{2}\frac{1-c}{z}\right)\right).
\end{equation}
with
\begin{equation}
  \zeta_2\lr{z} := z - c\sigma^2g_1^R\lr{z}. 
\end{equation}
This is a self-consistency equation for the block Stieltjes transform of $R$, $\bm{g}^R\lr{z}$, that depends on the noise variance $\sigma^2$, the aspect ratio $c$, and the standard Stieltjes transform of the signal covariance, $g_{YY^T}\lr{z}$.

Once this equation is solved, the singular vector overlaps can be obtained as well. We introduce notation for the real and imaginary parts of the block Stieltjes transform: $g_1^R\lr{\hat s}=h_1^R + if_1^R$, where we assume that the spectral density at $\hat s$ is finite. Then we insert this into \eqref{eq-Gaussian-R-transform} to get the real and imaginary parts of the block $\xr$-transform of $X$. After defining, for notational ease,
\begin{align}
  \nu\lr{z} := \re\zeta_2\lr{z} = z - c\sigma^2h_1^R\lr{z},
\end{align}
we can finally simplify the overlaps \eqref{eq-overlap-both-formulas} for the case of Gaussian noise:
\begin{subequations}\label{eq-overlaps-with-gaussian-noise}
\begin{align}
   \Phi_1\lr{\hat s, s} =&
  \frac{\nu\lr{\hat s}\xb\lr{\hat s} - c\sigma^2 \mathcal{A}\lr{\hat s, s}} {\lr{\mathcal{A}\lr{\hat s,s}}^2 + \lr{f_1^R\xb\lr{\hat s}}^2} \label{eq-left-overlap-with-gaussian-noise}\\
  \Phi_2\lr{\hat s, s} =&
  \frac{\lr{\nu\lr{\hat s} - \sigma^2\frac{1-c}{\hat s}} \xb\lr{\hat s} - c\sigma^2 \mathcal{A}\lr{\hat s, s}} {\lr{\mathcal{A}\lr{\hat s,s}}^2 + \lr{f_1^R\xb\lr{\hat s}}^2},
\end{align}
\end{subequations}

where we have
\begin{subequations}
\begin{align}
  \mathcal A\lr{\hat s, s} =& \nu\lr{\hat s} \lr{\nu\lr{\hat s} - \sigma^2\frac{1-c}{\hat s}} - \lr{s^2 + c^2\sigma^4\lr{f_1^R}^2}\\
  \mathcal B\lr{\hat s} =& 2c\sigma^2 \lr{\nu\lr{\hat s} - \sigma^2\frac{1-c}{2\hat s}}.
\end{align}

\end{subequations}

Formula \eqref{eq-left-overlap-with-gaussian-noise} is confirmed in Figure \ref{fig:Figure 3 - Sums of Gaussians}, which shows the left singular vector overlaps between data and signal, when the signal, $Y$, is Gaussian as well. The bottom colorplot can be thought of as an input-output map for the singular value structure under additive noise. It shows that a signal singular vector associated with a given singular value $s$ undergoes, loosely speaking, both ``diffusion'' and ``inflation'', aligning partially with data singular vectors across a range of singular values with a peak associated with larger singular values, $\hat s> s$. In the upper three panels we observe that individual overlaps are not self-averaging - a smooth overlap function emerges only when one averages either over many overlaps within a range of singular values, or over many instantiations.

We stress that these formulas for the overlap of data singular vectors with signal singular vectors do not depend directly on the unobserved signal $Y$. Rather, they depend only on the noise variance and the block Stieltjes transform, $g_1^R\lr{z}$, of the noisy data matrix, $R$. Furthermore, $g_1^R\lr{z}$ can be estimated empirically via kernel methods for the empirical spectral density and its Hilbert transform \cite{Ledoit2012, Ledoit2020, potters_bouchaud_2020}. This suggests that significant information about the structure of the unobserved extensive signal can be inferred from noisy empirical data, and this will lay the foundation for the optimal estimators derived below.

\subsection{SVD of the Extensive Spike Model}\label{subsec:SVD of extensive spike}

We now return to the spiked matrix model $R=Y+X$, with signal $Y=sU_1U_2^T$, where $s$ is a scalar, $U_a$ are $N_a\times K$ matrices with orthogonal columns. But now we assume the rank of the spike grows linearly with the number of rows at a fixed rank ratio, $b$, i.e. $K=bN_1$, while the aspect ratio $c=\nicefrac{N_1}{N_2}$ is fixed as before. We will assume the elements of the noise matrix $X$ are i.i.d. Gaussian: $X_{ij}\sim \xnorm\lr{0,\frac{1}{N_2}}$. In the following we first discuss the singular values, and then the singular vectors of the extensive-rank model. 

\subsubsection{Singular Value Spectrum \\ of the Extensive Spike Model}

$YY^T$ has $K$ eigenvalues equal to $s^2$ and $N_1-K$ zero eigenvalues. Its Stieltjes transform can therefore be found to be 
\begin{equation}
  g_{YY^T}\lr{z} = \frac{z+\left(b-1\right)s^2}{z\left(z-s^2\right)}.
\end{equation}
We can now make use of the self-consistency equation for $g_1^R\lr{z}$, \eqref{eq:Arbitrary_Signal_Gaussian_noise}. Momentarily writing $g$ in place of $g_1^R\lr{z}$ and simplifying, we find
\begin{equation}\label{eq-extensive-spike-Stieltjes-polynomial}
  \lrb{\lr{\zeta_2-\frac{1-c}{z}}g-1}\lrb{\lr{\zeta_2-\frac{1-c}{z}}\zeta_2-s^2}=bs^2,
\end{equation}
where we write $\zeta_2 = z-c\sigma^2 g$ as above. This is a quartic polynomial for $g=g_1^R\lr{z}$. We solve this numerically for $z$ near the real line in order to find the density of singular values of $R$ (see Appendix \ref{sec:Stieltjes-polynomial-Appendix} for the polynomial coefficients and details of numerical solution).

\begin{figure}
  \centering
  \includegraphics[height=13cm]{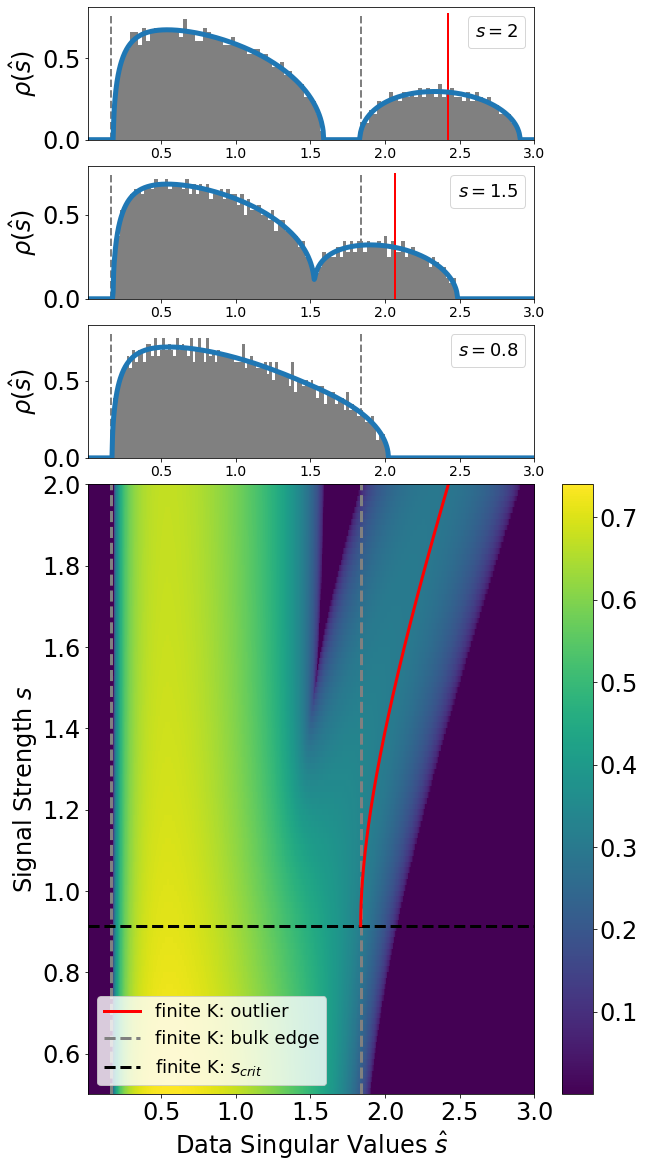}
  \caption{\textbf{Signal-Strength Transition in the SV Density of the Extensive Spike Model.} Each row of the bottom color map shows theoretical predictions for the singular value density of the extensive spike model, $\rho_1^R\lr{\hat s}$, corresponding to different signal strengths $s$ (along y-axis) at a fixed rank ratio of $b=\nicefrac{K}{N_1} = 0.25$. In all panels in this figure, the aspect ratio is fixed to $c=\nicefrac{N_1}{N_2}=0.7$. Features of the finite-rank spike model are shown as lines for comparison. The horizontal black dashed line indicates the threshold signal strength $s_{crit}$ above which the finite-rank model acquires an outlier singular value. The red curve indicates the position of this outlier singular value. The vertical grey dashed lines indicate the edges of the bulk spectrum of the finite-rank model. The top $3$ panels, corresponding to horizontal slices of the color maps, plot the singular value density at $3$ different signal strengths $s=0.8,1.5$ and $2$. Solid blue curves indicate theoretical predictions from numerically solving \eqref{eq-extensive-spike-Stieltjes-polynomial}, while grey histograms indicate the spectral density from a single realization with $N_2=2000$. For comparison, the red vertical spike indicates the position of outlier singular value in the finite-rank theory, while the grey dashed spike indicates the edge of the noise bulk in this theory. Together these panels demonstrate that as $s$ increases, the singular value density first undergoes a crossover from a unimodal to a bimodal regime, and then a phase transition from a connected to a disconnected phase. }
  \label{fig:Figure 4 -- sv density signal strength phase diagram}
\end{figure}

For strong signal $s$, the spectrum in the extensive case differs from the finite rank case most clearly in that singular values reflecting the signal do not concentrate at a single data singular value. Rather, (see Figure \ref{fig:Figure 4 -- sv density signal strength phase diagram} top) for sufficiently strong signal $s$, the presence of noise blurs the signal singular values into a continuous bulk that is disconnected from the noise bulk. This signal bulk appears near the single outlier predicted by the finite-rank theory, but has significant spread.

At very weak signals $s$ there is a single, unimodal bulk spectrum, just as in the finite-rank setting, but in contrast, these weak signals make their presence felt by extending the leading edge of the bulk beyond the edge of the spectrum predicted by the finite-rank theory, {\it even} when the signal strength $s$ is below the critical signal strength $s_{crit}$ predicted by finite-rank model (Figure \ref{fig:Figure 4 -- sv density signal strength phase diagram} 3rd panel).

At intermediate signal strength $s$, the singular value distribution exhibits a connected bimodal regime not present in the finite-rank model (Figure \ref{fig:Figure 4 -- sv density signal strength phase diagram} 2nd panel).

Thus, as $s$ increases, we see two qualitative changes: first a crossover from a single unimodal bulk to a single bimodal bulk, and then from one connected bulk to two disconnected bulks. This final splitting of the signal bulk from the noise bulk is a phase transition as the block Stieltjes transform goes from having a single branch cut to two disjoint branch cuts. This transition happens at significantly larger signal $s$ than the signal-detectability phase transition in the finite-rank regime (Figure \ref{fig:Figure 4 -- sv density signal strength phase diagram} bottom). 

\begin{figure*}
  \centering
  \includegraphics{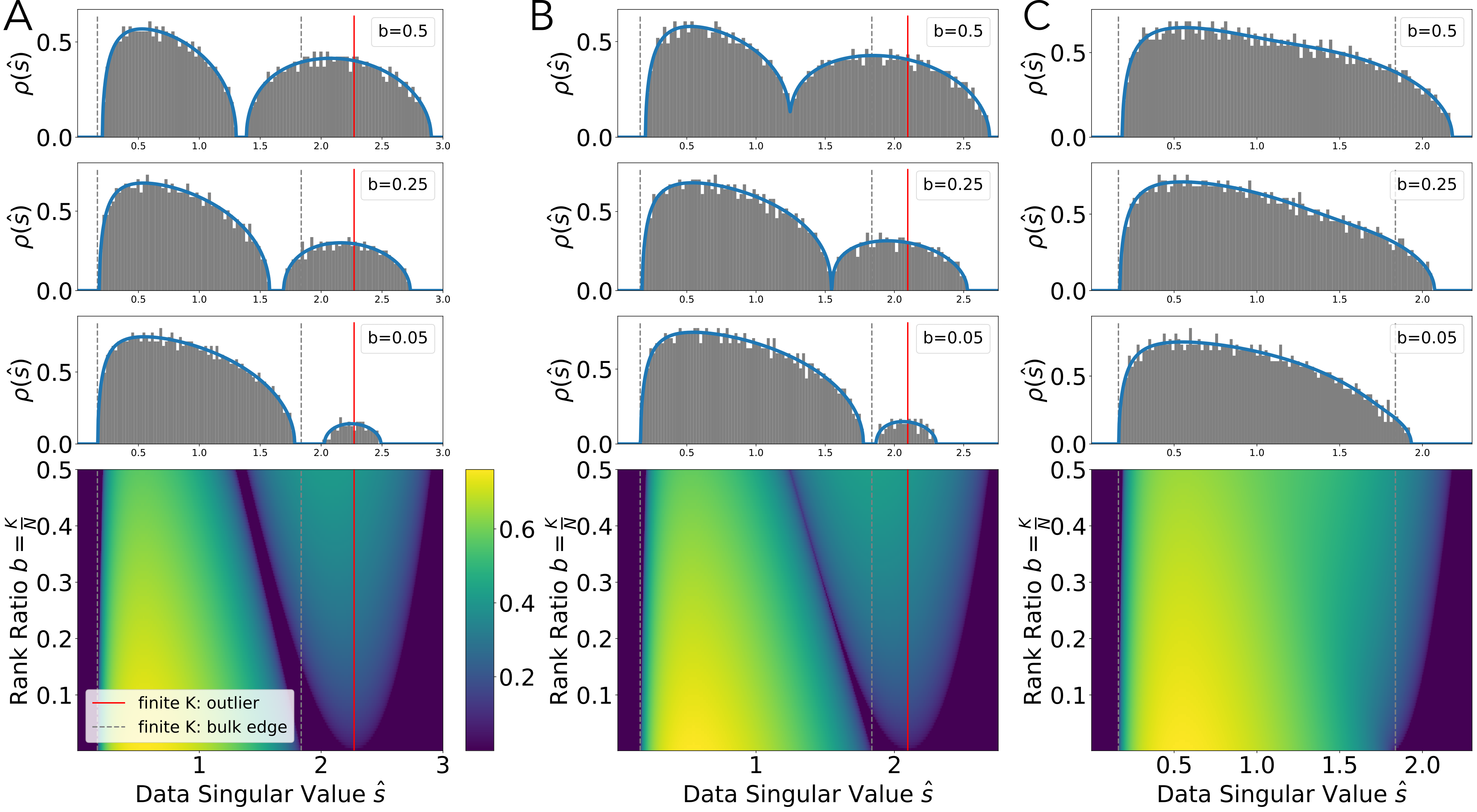}
  \caption{\textbf{Rank-Ratio Transitions in the SV Density of the Extensive Spike Model.} Each row of the bottom color maps show theoretical predictions for the singular value density $\rho_1^R\lr{\hat s}$ corresponding to different rank ratios $b$ (y-axis) at a fixed signal strength, and all color maps have same color scale. Top $3$ panels indicate matching theory (blue curves) and empirics of a single realization (grey histograms) for $3$ rank ratios $b=0.05,0.25,0.5$, and the aspect ratio is fixed to $c=\nicefrac{N_1}{N_2}=0.7$ with $N_2=2000$. Comparisons to the finite-rank theory are shown using the same conventions as in Fig \ref{fig:Figure 4 -- sv density signal strength phase diagram}. \textbf{A.} Results for $s=1.8$, illustrating that for sufficiently strong signal the singular value density remains in the disconnected phase for all values values of $b$. \textbf{B.} Results for $s=1.55$, illustrating that for intermediate signal strengths the density undergoes a transition from disconnected to connected as the rank ratio $b$ increases. \textbf{C.} Results for $s=0.9$, illustrating that for subthreshold signals the density remains connected for all $b$.  $s_{crit}=c^{\nicefrac{1}{4}}\approx 0.915$ throughout.}
  \label{fig:Fig5_Final Rank-Ratio SV Density Phase Transition}
\end{figure*}

In the limit of low rank (small $b$) the spectrum approaches the finite-rank theory as expected (Figure \hyperref[fig:Fig_S1_lowrank_lim]{S1}). Interestingly, we find that as a function of rank ratio $b$, there are three distinct regimes. For sufficiently strong signals (Figure \ref{fig:Fig5_Final Rank-Ratio SV Density Phase Transition}A), the signal bulk remains disjoint from the noise bulk for all $b$. For intermediate signals (Figure \ref{fig:Fig5_Final Rank-Ratio SV Density Phase Transition}B), the two bulks merge but the spectrum remains bimodal for all $b$. Finally, for weak signals (Figure \ref{fig:Fig5_Final Rank-Ratio SV Density Phase Transition}C), there is a single connected bulk for all $b$.

\subsubsection{Singular Vector Subspace Overlap \\ in the Extensive Spike Model}
We now turn to the singular vectors of the extensive spike model. For simplicity we focus on the left-singular vectors. Since the $K$ non-zero singular values of the signal are degenerate, the only meaningful overlap to study is a subspace overlap, or the projection of the data singular vectors, $\hat{\bm{u}}_{\bm{1}k}$, onto the entire subspace defined by $U_1$. Therefore we compute
\begin{equation}
  \left\Vert \hat{\bm{u}}_{\bm{1}k}^TU_1\right\Vert^2=\sum_{m=1}^K\lr{\hat{\bm{u}}_{\bm{1}k}^T\bm{u}_{\bm{1}m}}^2.
\end{equation}
Since this is an extensive sum, we expect that it is self-averaging, and should be well-predicted by $ b\Phi_1\lr{\hat s_k, s}$, where $\bm\Phi$ is defined in \eqref{eq-overlaps-with-gaussian-noise}. 

After solving \eqref{eq-extensive-spike-Stieltjes-polynomial} for the block Stieltjes transform of $R$, we insert the result in \eqref{eq-overlaps-with-gaussian-noise} to find $\Phi_1\lr{\hat s,s}$. In Figure \hyperref[fig:Fig_S2_extensive_spike]{S2} we return to the simulation results presented in Figure \ref{fig:Figure 2 finite-rank theory fails} and show that the extensive-rank theory predicts both the leading outlier singular value and the subspace overlap of the corresponding singular vector, even when the finite-rank theory fails.

\begin{figure}
  \centering  
  \includegraphics[width=8.6cm]{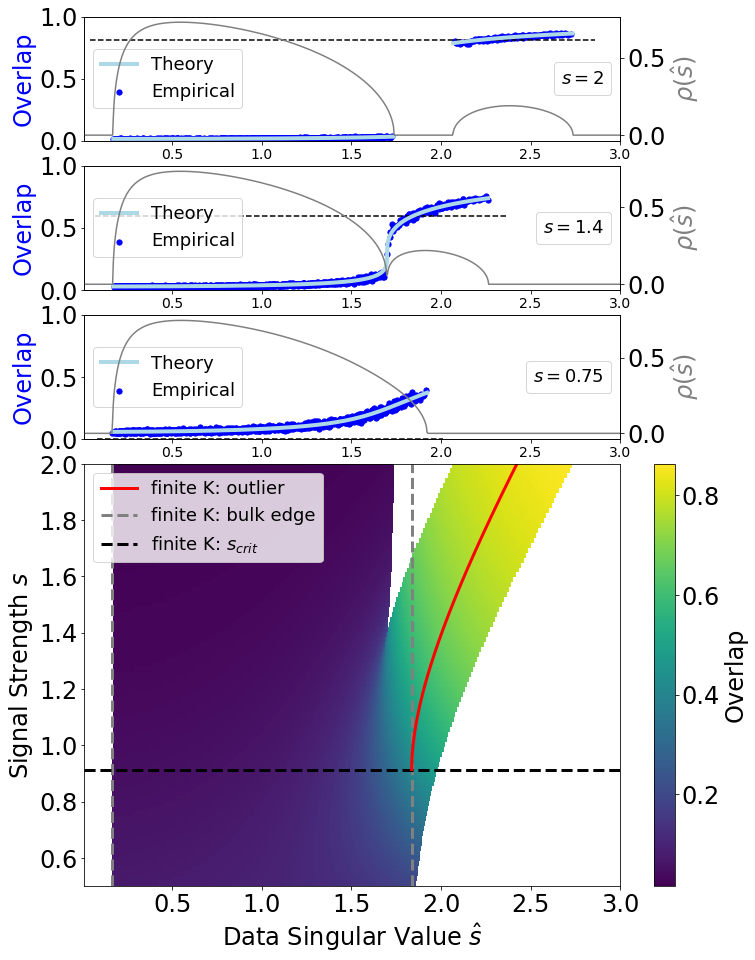}
  \caption{\textbf{Singular Vector Overlaps in the Extensive Spike Model.} Each row of the bottom color plot shows the theoretical prediction for the overlap of a left singular vector with singular value $\hat s$ of the data matrix $R=Y+X$, with the entire $K$ dimensional signal subspace of $Y$ (i.e. the squared norm of the projection of the singular vector onto this subspace). The prediction is given by $b\Phi_1\lr{\hat s,s}$, using \eqref{eq-overlaps-with-gaussian-noise} after numerically solving for $g_1^R\lr{\hat s}$ from \eqref{eq-extensive-spike-Stieltjes-polynomial}. Different rows along the y-axis correspond to different signal strengths $s$ for $Y$. Comparisons to the finite-rank theory are shown using the same conventions as in Fig. \ref{fig:Figure 4 -- sv density signal strength phase diagram}. The top $3$ panels show horizontal slices for $s=0.75,1.4,2.0$. Solid grey curves indicate the singular value density of the data matrix $R$ in the extensive-rank model. Blue dots indicate numerical calculations for the overlap for a single realization with $N_2=2000$. Solid light blue lines through the blue dots indicate matching theoretical predictions for this overlap. For comparison, the horizontal dashed line indicates the overlap predicted by the finite-rank theory (which depends only on $s$ and not $\hat s$). The 3rd panel indicates that the signal subspace of $Y$ is detectable in the top data singular vectors of $R$, {\it even} at small signal strengths $s$ {\it below} the transition in the singular value density of $R$ from unimodal to bimodal. The aspect ratio is $c=\nicefrac{N_1}{N_2}=0.7$, while the rank ratio for $Y$ is fixed at $b=\nicefrac{K}{N_1}=0.1$.}
  \label{fig:Figure 6 - sv overlap extensive spike}
\end{figure}

In Figure \ref{fig:Figure 6 - sv overlap extensive spike} we explore the phase diagram of the extensive-rank model and successfully confirm the predictions of the extensive-rank theory for singular vector overlaps by comparing these predictions to numerical simulations.

For strong signal $s$ (Fig \ref{fig:Figure 6 - sv overlap extensive spike} top panel), the overlap of the data singular vectors with the true signal subspace is reasonably approximated by the finite-rank theory \cite{Benaych-Georges2012}. However, for moderate signals (Fig \ref{fig:Figure 6 - sv overlap extensive spike} second panel) the data singular vectors interact, competing for the signal subspace. Singular vectors associated with the leading edge of the signal bulk have higher subspace overlap with the signal, while those at the lower edge overlap less. Perhaps most intriguingly, even for weak signals below the finite-rank phase transition at $s=s_{crit}$ the top data singular vectors still overlap significantly with the signal subspace (Fig \ref{fig:Figure 6 - sv overlap extensive spike} third panel). Note, this overlap is nontrivial and $O(1)$ {\it even} when the singular value spectrum of the data is in the unimodal bulk regime.

We observe that the extensive spike model exhibits a singular value inflation in its {\it singular-vector overlaps}. Not only are data singular values larger than the corresponding signal singular values, just as in the finite-rank model, but also the singular-vector overlap peaks at the upper edge of the data singular values.

\begin{figure}
  \centering
  \includegraphics[height=5.8cm, width=8.5cm]{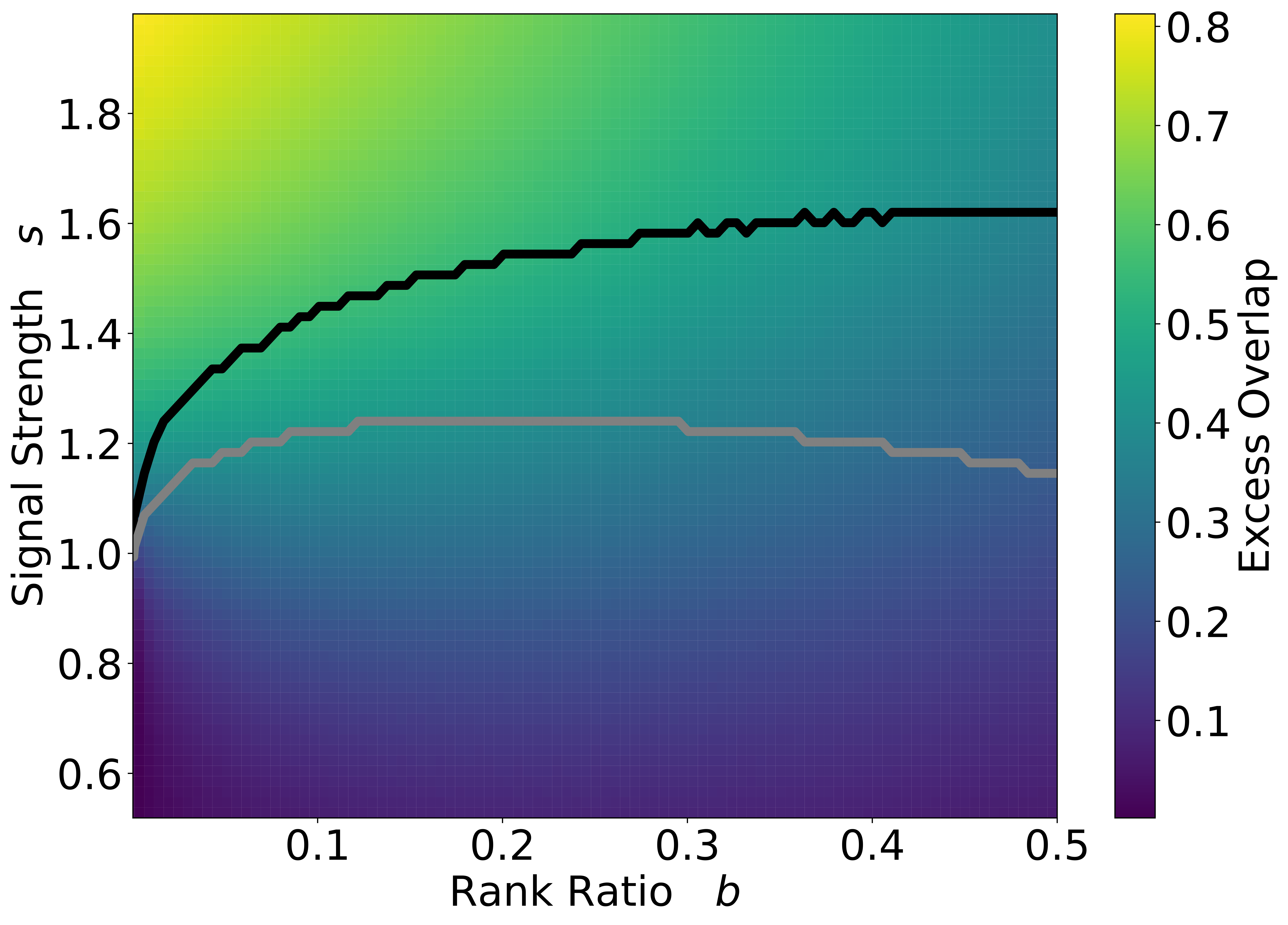}
  \caption{\textbf{Singular Vector Overlaps Disregard Singular Value Phases} Two-dimensional phase diagram shows the average ``excess' subspace overlap \eqref{eq-excess-overlap} of the top $b$ fraction of data singular vectors with a signal of strength $s$ (y-axis) and rank ratio $b$ (x-axis). The (lower) grey line separates the unimodal and bimodal regimes of the SV spectrum, and the (upper) black line separates the connected phase from the disconnected phase. The singular vector overlap does not respect the boundaries of the SV spectrum. The signal impacts the data via significant overlaps with the signal subspace well below the boundary between unimodal and bimodal regimes. Aspect ratio $c=0.7$}
  \label{fig:Fig7 2D phase diagram}
\end{figure}

Figure \ref{fig:Fig7 2D phase diagram} summarizes the results of this section with a two-dimensional phase diagram in the signal-strength vs rank ($s$-$b$) plane. It shows the boundaries between three regimes of the singular value spectrum: unimodal, bimodal, and disconnected. Additionally, the color map shows the average {\it excess} signal subspace overlap of the singular vectors associated with the top $b$ fraction of singular values. Since by chance, any random vector is expected to have an overlap $b$ with the signal subspace, we compute the excess overlap as $\tilde{\Phi}\lr{\hat s,s}=b\lr{\Phi\lr{\hat s,s} -1}$. We then average the excess overlap across the singular vectors associated with the top $b$ singular values, that is:
\begin{equation}\label{eq-excess-overlap}
\int_{t}^\infty \tilde{\Phi}\lr{\hat s,s}\rho_1^R\lr{\hat s}\rm d\hat s,
\end{equation}
where $t$ is given by $b=\int_{t}^\infty\rho_1^R\lr{\hat s}\rm d\hat s$.

Importantly, the figure demonstrates that in contrast to the finite-rank setting, the transitions in the data singular value spectrum of the data do {\it not} coincide with the detectability of the signal. Rather, the alignment of the data singular vectors with the signal subspace is a smooth function of both signal strength $s$ and rank ratio $b$, and nonzero excess overlap can occur even in the unimodal regime. 

\section{Optimal Rotationally Invariant Estimators }\label{sec:optimal rie}

We now consider two estimation problems given noisy observations, $R=Y+X$: 1) denoising $R$ in order to optimally reconstruct $Y$, and 2) estimation of the true signal covariance, $C=YY^T$. We focus on the case where both signal $Y$ and noise $X$ rotationally invariant ($P_Y(M)=P_Y(O_1 M O_2)$ for arbitrary orthogonal matrices $O_1,O_2$, and similarly for $X$). In this setting it is natural to consider {\it rotationally invariant} estimators $F$ that transform consistently with rotations of the data: $F\left(O_1 R O_2\right)=O_1 F(R) O_2$ \cite{Takemura1984,Stein1986}. Such $F$ can only alter the singular values of $R$ while leaving the singular vectors unchanged. More generally, when $Y$ is not rotationally invariant, our results yield the best estimator that only modifies singular values of $R$. 

Our problem thus reduces to determining optimal shrinkage functions for the singular values. In the finite-rank case, distinct singular values and their associated singular vectors of $Y$ respond independently to noise, so the optimal shrinkage of $\hat{s}$ depends only on $\hat{s}$ \cite{Shabalin2013,Gavish2017,Donoho2018}. As we show below, this is no longer the case in the extensive-rank regime. The optimal shrinkage for each singular value generally depends on the entire data singular value spectrum.

\subsection{Denoising Rectangular Data}\label{subsec:denoising}

We first derive a minimal mean-square error (MMSE) denoiser to reconstruct the rotationally invariant signal, $Y$, from the noisy data, $R$. Under the assumption of rotational invariance, the denoised matrix is constrained to have the same singular vectors as the data $R$, and thus takes the form $\tilde Y = \hat U_1 \phi\lr{\hat S}\hat U_2^T$. The MSE can be written
\begin{align}
  \mathcal {E} &= \frac{1}{N_1N_2}\rm{Tr} \lr{Y-\tilde Y}\lr{Y-\tilde Y}^T \nonumber \\
  &= \frac{1}{N_1N_2} \sum_m s_m^2 + \phi^2\lr{\hat s_m} - 2\phi\lr{\hat s_m}\hat{\bm{u}}_{\bm{1}m}^T Y \hat{\bm{u}}_{\bm{2}m}.
\end{align}
Minimizing with respect to $\phi(\hat{s}_m)$ gives the optimal shrinkage function:
\begin{equation}
  \phi^*\lr{\hat s_m} = \hat{\bm{u}}_{\bm{1}m}^T Y \hat{\bm{u}}_{\bm{2}m},
\end{equation}
which appears to require knowledge of the very matrix being estimated, namely $Y$. However, in the large size limit it is possible to estimate $\phi^*\lr{\hat s_m}$ via the resolvent $G^R(z)$.
We first write
\begin{align}  \mathrm{Tr}\lrb{\bm{Y}\bm{G}^R\lr{z}}_{11} =& \mathrm{Tr}\lrb{YR^TG_{RR^T}\lr{z^2}} \nonumber \\
  =&\sum_{l}\frac{\hat s_l }{z^2 - \hat s_l^2} \hat{\bm{u}}_{\bm{1}l}^T Y \hat{\bm{u}}_{\bm{2}l}.
\end{align}
As $z$ is brought toward the singular value $\hat{s}_m$ the sum is increasingly dominated by the contribution from $\hat{\bm{u}}_{\bm{1m}}^T Y \hat{\bm{u}}_{\bm{2m}}=\phi^*\lr{\hat s_m}$. We find
\begin{equation}
\label{eq-phi-calculation}
  \phi^*\lr{\hat s} = \frac{2}{\pi\rho^R_1\lr{\hat s}} 
  \xlim{\eta}{0}\im\lrb{\tau_1\lrb{\bm Y \bm G^R\lr{\hat s - i\eta}}}.
\end{equation}

We next apply the subordination relation \eqref{eq-subordination-main}, yielding a product of $Y$ with a $Y$-resolvent, whose trace is readily found:
\begin{equation}
  \tau_1\lrb{\bm Y \bm G^R\lr{z}}=\tau_1\lrb{\bm Y \bm G^Y\lr{\bm{\zeta}}}= \zeta_1 g^Y_1(\bm{\zeta}) - 1,
\end{equation}
where $\zeta_a\lr{z} = z - \xr _a^X\lr{\bm{g}^R\lr{z}}$, and we have used the identity $\tau\lrb{CG_C\lr{z}}=zg_C\lr{z}-1$ for arbitrary symmetric $C$.

Since $g_1^Y\lr{\bm{\zeta}}=g_1^R\lr{z}$ \eqref{eq-subordination-stieltjes}, we obtain
\begin{equation}
\label{eq-rie-estimator-gen}
  \phi^*\lr{\hat s} = \frac{2}{\pi\rho_1^R\lr{\hat s}} 
  \xlim{\eta}{0}\im\lrb{\zeta_1\lr{\hat s-i\eta}g^R_1\lr{\hat s-i\eta}},
\end{equation}
which depends only on the block Stieltjes transform of the empirical data matrix, $R$, and the block $\xr$-transform of the noise, $X$. Importantly, the dependence on the unknown signal $Y$ is gone, making this formula amenable to practical applications, at least when the noise distribution of $X$ is known. 

For i.i.d. Gaussian noise with known variance, $\frac{\sigma^2}{N_2}$, we have $\xr^X_1\lr{\bm g}=\sigma^2g_2$ and the general relation $g_2\lr{z}=cg_1\lr{z}+\frac{1-c}{z}$, so \eqref{eq-rie-estimator-gen} simplifies considerably. Writing the real and imaginary parts, $g^R_1\lr{z} = h^R_1\lr{z} + if^R_1\lr{z}$, we obtain the following simple expression depending only on the variance of the noise, and the Hilbert transform of the observed data spectral density:
\begin{equation}
  \label{eq-gaussian-shrinkage}
  \phi^*\lr{\hat s} = \hat s - 2c\sigma^2 h^R_1\lr{\hat s} - \sigma^2 \frac{1-c}{\hat s}.
\end{equation}
This expression for the Gaussian case was derived previously in \cite{Troiani2022}.

\begin{figure*}
  \centering
  \includegraphics{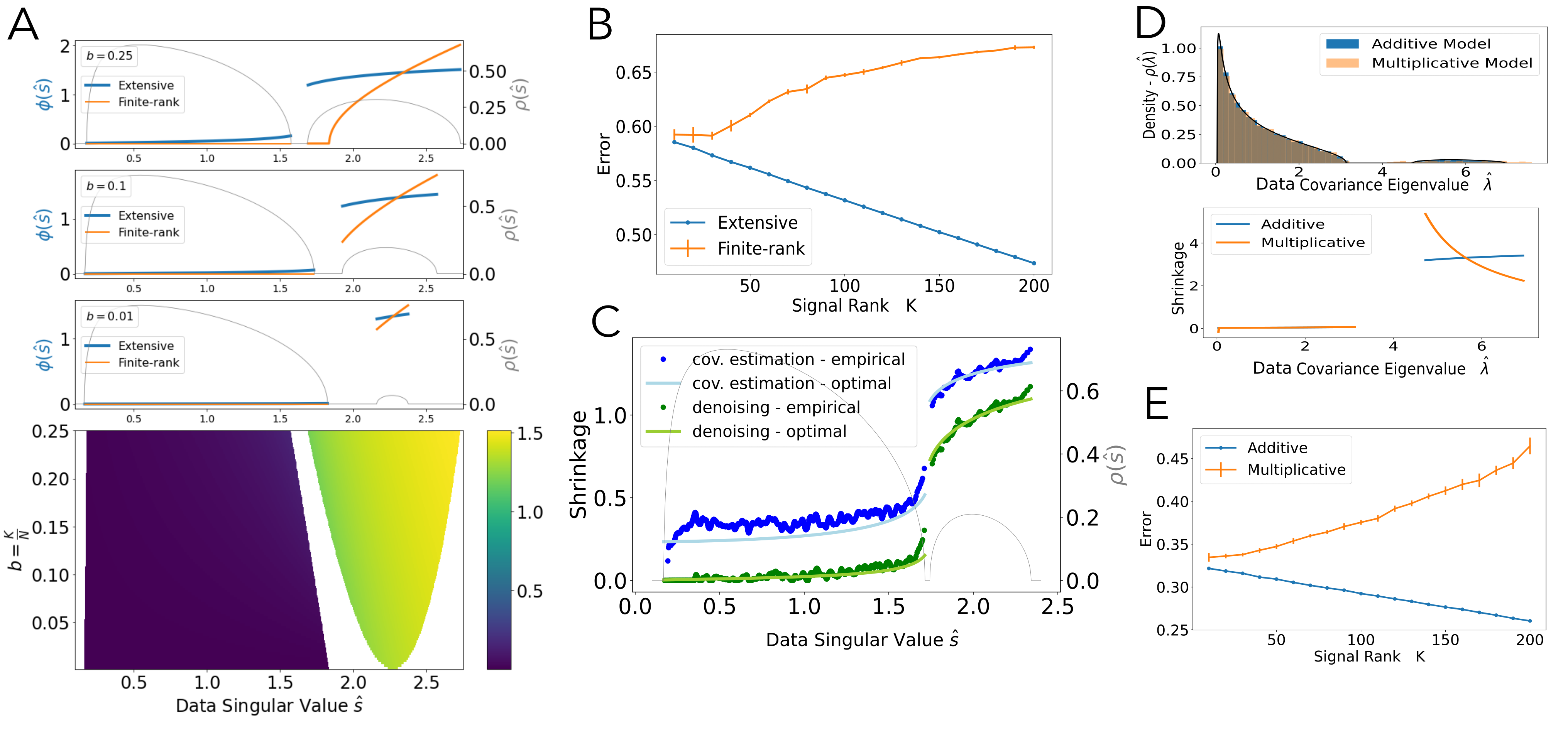}
  \caption{\textbf{Optimal Denoising of Extensive Spikes. A} Each row of the bottom color map shows the optimal shrinkage function $\phi^*\lr{\hat s}$ \eqref{eq-gaussian-shrinkage} for denoising data from the extensive spike model. Different rows on the y-axis correspond to different rank ratios $b=\nicefrac{K}{N_1}$ of the signal $Y$, while the signal strength $s$ of $Y$ is fixed at $s=1.8$ and the aspect ratio is fixed at $c=0.7$. The top 3 panels show horizontal slices with $b=0.25,0,1,0.01$. Blue (darker) curves indicate the optimal shrinkage function for the extensive-rank model while orange (lighter) curves indicate the optimal shrinkage function for the finite-rank model \cite{Gavish2017} Eq (7) (which does not depend on $b$). These panels indicate that the optimal shrinkage function for the extensive-rank model balances singular values more than that of the finite-rank model by more (less) aggressively shrinking larger (smaller) singular values. \textbf{B.} Comparison of mean-square error in rectangular data denoising of $K$ spikes, as a function of $K$ for fixed signal strength $s=2$, using the optimal shrinkage function for the finite-rank model (orange) versus that of the extensive spike model (blue). Even at small spike numbers of $K=10$ for $N_1 = 1000$ by $N_2 = 500$ sized data matrices, the extensive denoiser outperforms the finite-rank denoiser, and at larger $K$ the extensive (finite-rank) denoiser gets better (worse). \textbf{C.} Empirical shrinkage, and comparison of optimal shrinkage function for two different errors: in blue (darker) denoising the rectangular signal matrix, $\phi^*\lr{\hat s}$ \eqref{eq-gaussian-shrinkage}, and in green (lighter) estimating the $N_1\times N_1$ signal covariance matrix $\sqrt{\psi^*\lr{\hat s}}$ \eqref{eq-rie-cov-gaussian}, for signal strength $s=1.5$ and rank ratio $b=0.1$, with aspect ratio $c=0.7$. Lighter curves show the theoretical optima found using Eq \eqref{eq-extensive-spike-Stieltjes-polynomial} Darker dots show empirical shrinkage obtained via kernel estimation (see Appendix \ref{sec:Kernel-estimates-empirical-Appendix}) of the block Stieltjes transform from the data singular values with $N_2=2000$. \textbf{D.} Comparison between multiplicative model (spiked covariance) and additive model (spiked rectangular model) with $K=50$. \textit{Top:} Eigenvalue spectra of data covariance ($RR^T$) for multiplicative model and additive model. \textit{Bottom:} Optimal shrinkage for covariance estimation under the wrong model. Data spectrum generated by additive model, shrinkage function of multiplicative model \cite{Ledoit2011} Eq (13) vs the correct, additive model. \textbf{E.} Mean-square error in covariance estimation as a function of $K$ using multiplicative model (orange) vs correct, additive model (blue). Throughout D and E, $s=2$ with $N_1=1000$ and $N_2=1500$, and multiplicative model is displayed in orange (lighter) and additive model is displayed in blue (darker).}
 \label{fig:Figure_8_Optimal_Estimation}
\end{figure*}

\begin{figure}
\renewcommand{\figurename}{FIG. S1}
\renewcommand{\thefigure}{}
  \centering
  \includegraphics[width=8.6cm]{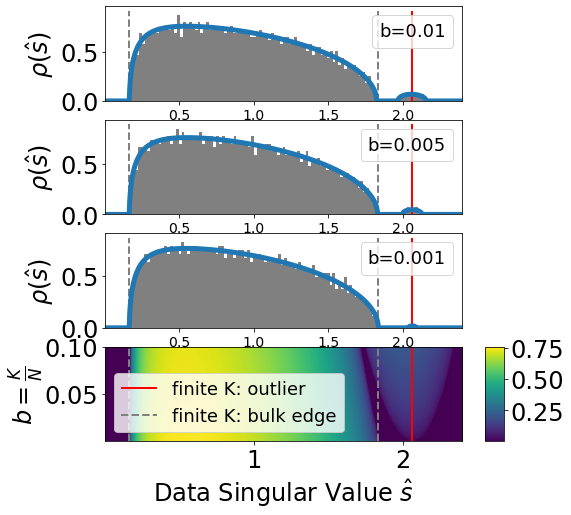}
  \caption{\textbf{Singular Value Density of Extensive Spike Model in the Low-Rank Limit.} Color plot shows singular value density of extensive spike model with aspect ratio $c=\nicefrac{N_1}{N_2}$ and signal strength $s=1.5$. The rank ratio, $b=\nicefrac{K}{N_1}$, varies along the y-axis. Top 3 panels show horizontal slices for $b=0.001,0.005,0.01$ together with empirical histograms of individual model instantiations with $N_2=2000$. The extensive-rank theory converges to the finite-rank theory as $b$ gets small.}
  \label{fig:Fig_S1_lowrank_lim}
\end{figure}

\begin{figure}
\renewcommand{\figurename}{FIG. S2}
\renewcommand{\thefigure}{}
  \centering
  \includegraphics[width=8.6cm]{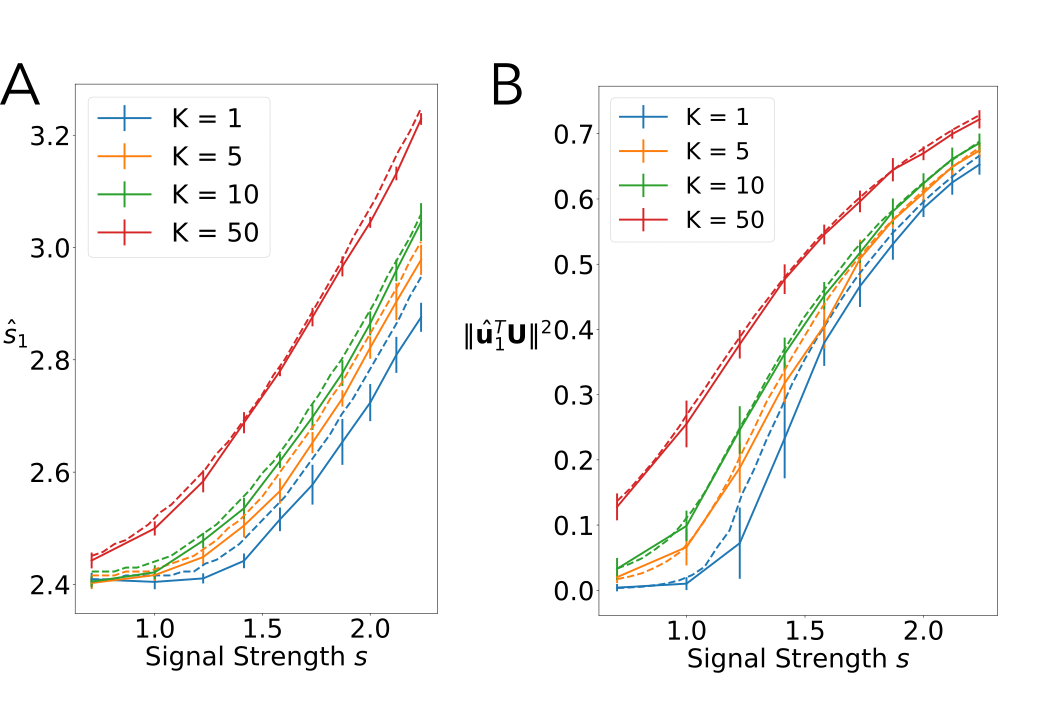}
  \caption{\textbf{Extensive-Rank Theory Captures The Singular Value Structure of The Spiked Rectangular Matrix Model.} Compare Figure \ref{fig:Figure 2 finite-rank theory fails}. \textbf{A.} Leading data singular value, $\hat s_1$, as a function of signal singular value, $s_1$, for various ranked spikes. \textbf{B.} Projection of leading data left singular vector, $\mathbf{u}_1$ , on the $K$-dimensional left singular space of the signal. Dashed lines show extensive-rank theory. The two panels match the Figure \ref{fig:Figure 2 finite-rank theory fails}(A and B), with $N_1=1000$ and $N_2=500$ and numerical results presented as mean and standard deviation over $10$ instantiations for each value of $b$ and $s$. This figure shows that the extensive-rank theory captures the deviations from finite-rank theory at finite $N_1$ and $N_2$.}
  \label{fig:Fig_S2_extensive_spike}
\end{figure}

Figure \ref{fig:Figure_8_Optimal_Estimation} compares \eqref{eq-gaussian-shrinkage} to the optimal shrinkage found based on the finite-rank theory \cite{Gavish2017}. The extensive-rank formulas recover many more significant singular values (Figure \ref{fig:Figure_8_Optimal_Estimation}A). Moreover the mean-square error of $Y^*=\hat{U_1}\phi^*\lr{\hat S}\hat{U_2}$ is superior to that of the finite-rank denoiser, steadily improving as a function of the signal rank, while the finite-rank denoiser worsens (Figure \ref{fig:Figure_8_Optimal_Estimation}B). In fact, for our simulations with $N_1=1000$ and $N_2=500$, the extensive-rank denoiser out performed the finite-rank denoiser for all $K>5$, across the range of signal strengths tested. Finally, given an estimate of the noise variance $\sigma^2$, we are able to numerically estimate $g_1^R\lr{\hat s}$ with kernel methods (Appendix \ref{sec:Kernel-estimates-empirical-Appendix}) and compute an empirical shrinkage function that is very close to the theoretical optimum (Figure \ref{fig:Figure_8_Optimal_Estimation}C). 

\subsection{Estimating the Signal Covariance}\label{subsec:cov estimation}

We now derive an MMSE-optimal rotationally invariant estimator for the signal covariance, $C=YY^T$. Just as in \cite{Ledoit2011, Bun2016}, and similarly to our results in the previous section, the optimal estimator is given by $C^*=\hat U_1, \psi^*\lr{\hat S}\hat U_1^T$, where:
\begin{equation}
  \psi^*\lr{\hat s_l} = \hat{\bm{u}}_{\bm{1}l}^TC \hat{\bm{u}}_{\bm{1}l}.
\end{equation}
We observe that the top-left block of the square of the Hermitianization $\bm Y$ is given by $C$, and so
\begin{equation}
  \lrb{\bm{Y}^2\bm{G}^R\lr{z}}_{11} = \sum_{l} \frac{z}{z^2-\hat{s}_l^2}  \hat{\bm{u}}_{\bm{1}l}^T C \hat{\bm{u}}_{\bm{1}l}.
\end{equation}
Thus, we can calculate the optimal shrinkage function by the inversion relation \eqref{eq-inversion-relation}:
\begin{equation}
  \psi^*\lr{\hat s} = \frac{2}{\pi\rho^R_1\lr{\hat s}}\xlim{\eta}{0} \im \lrb{\tau_1\lrb{\bm{Y}^2\bm{G}^R\lr{\hat s-i\eta}}}.
\end{equation}
Now, we apply the subordination relation, $\bm{G}^R\lr{z} = \bm{G}^Y\lr{\bm{\zeta}\lr{z}}$ with $\zeta_a = z - \xr _a^X\lr{\bm{g}^R\lr{z}}$, which gives $\bm{Y}^2\bm{G}^R\lr{z} = \bm{Y}^2\bm{G}^Y\lr{\bm{\zeta}}$, which has top-left block $\zeta_2YY^TG_{YY^T}\lr{\zeta_1\zeta_2}$. 

Again, using the identity $\tau\lrb{CG_C\lr{z}}=zg_C\lr{z}-1$ for arbitrary symmetric $C$, we have
\begin{equation}\tau_{1}\left[\bm{Y}^{2}\bm{G}^{R}\left(z\right)\right]=\zeta_{2}\left(\zeta_{1}g^{Y}_{1}\left(\zeta_{1}\zeta_{2}\right)-1\right).
\end{equation}
We therefore conclude for general noise matrix, $X$:
\begin{align}\label{eq-rie-estimator-cov-gen}
  \psi^*\lr{\hat s} = \frac{2}{\pi\rho^R_1\lr{\hat s}} 
  \xlim{\eta}{0}\im &[\zeta_2\lr{\hat s-i\eta} \\
  &\lr{\zeta_1\lr{\hat s - i\eta} g^R_1\lr{\hat s-i\eta} - 1}] \nonumber.
\end{align}
Once again, for i.i.d. Gaussian noise with known variance, $\frac{\sigma^2}{N_2}$, our estimator \eqref{eq-rie-estimator-cov-gen} simplifies considerably. Using the optimal shrinkage function found above for rectangular denoising, $\phi^*\lr{\hat s} = \hat s - 2c\sigma^2 h^R_1 - \sigma^2 \frac{1-c}{\hat s}$, where $h^R_1$ is the real part of $g_1^R(\hat s)$, we finally obtain
\begin{align}\label{eq-rie-cov-gaussian}
  \psi\left(\hat s\right) =& \phi\lr{\hat s}\lr{\phi\lr{\hat s} + \sigma^2\frac{1-c}{\hat s}} \\
  &- c\sigma^2\lrb{c\sigma^2\left\vert g_1^R\right\vert^2 - 1} \nonumber.
\end{align}

Just as in optimal data denoising, we find that given an estimate of the noise variance, $\sigma^2$, the optimal shrinkage for covariance estimation depends only on the spectral density of $R$ and its Hilbert transform, which can be estimated directly from data.

In Figure \ref{fig:Figure_8_Optimal_Estimation} we show the optimal shrinkage function for the extensive spike model, and demonstrate that it can be approximated given only an estimate of the noise variance and the empirical data matrix, $R$ (Figure \ref{fig:Figure_8_Optimal_Estimation}C). We find that the optimal singular value shrinkage of singular values derived for covariance estimation \eqref{eq-rie-cov-gaussian}, $\sqrt{\psi\lr{\hat s}}$, is substantially different than $\phi\lr{\hat s}$ \eqref{eq-gaussian-shrinkage} obtained for denoising the rectangular signal (Figure \ref{fig:Figure_8_Optimal_Estimation}C). The denoising shrinkage suppresses the noise more aggressively, but suppresses the signal singular values more as well.

Finally, we compare the shrinkage obtained from assuming a multiplicative form of noise instead of the additive spiked rectangular model studied here. In the finite-rank regime, the spiked rectangular model can be instead modeled as a multiplicative model with data arising form a spiked covariance. Concretely, the data in the multiplicative model is generated as $R_{mult} = \sqrt{C_{mult}}X$, i.e. each column is sampled from a spiked covariance: $C_{mult} = YY^T + I$. In the finite-rank regime, with Gaussian noise, the two models yield identical spectra and covariance-eigenvector overlaps. The optimal shrinkage for covariance estimation for the multiplicative model for arbitrary $C_{mult}$ has previously been reported (\cite{Ledoit2011} Eq (13) for Gaussian noise and \cite{ Bun2016} Eq IV.8 for more general noise), and here we consider the impact of employing the multiplicative shrinkage formula on data generated from the additive spiked rectangular model. We observe (Figure \ref{fig:Figure_8_Optimal_Estimation}D Top) that for small rank-ratio ($b=0.05$) the two models give fairly similar eigenvalue distributions. Nevertheless, applying the optimal multiplicative shrinkage on the additive model data gives poor results: the shrinkage obtained is non-monotonic in the data eigenvalue (Figure \ref{fig:Figure_8_Optimal_Estimation}D Bottom). Furthermore, the mean-square error in covariance estimation obtained with the multiplicative shrinkage worsens as a function of rank (Figure \ref{fig:Figure_8_Optimal_Estimation}E).

\section{Discussion}

While one approach to estimation depends on prior information about the structure of the signal (such as sparsity of singular vectors for example), we have followed a line of work on rotationally invariant estimation that assumes there is no special basis for either the signal or the noise \cite{Takemura1984,Stein1986}. In this approach, knowledge of the expected deformation of the singular value decomposition (SVD) of the data due to noise allows for the explicit calculation of optimal estimators.

In the case of finite-rank signals, where the impact of additive noise on singular values and vectors is known \cite{Loubaton2011, Benaych-Georges2012}, formulas for optimal shrinkage for both denoising \cite{Shabalin2013,Gavish2014,Gavish2017} and covariance estimation \cite{Donoho2018} have been found. For extensive-rank signals, however, while formulas for the singular value spectrum of the free sum of rectangular matrices are known \cite{Benaych-Georges2009,Speicher2011,Mingo2017}, there are no prior results for the singular vectors of sums of generic rectangular matrices (though see \cite{Pourkamali2023} for contemporaneous results).

Even in the setting of square, Hermitian matrices, results on eigenvectors of sums are relatively new \cite{Ledoit2011, Allez2014}. Recent work derived a subordination relation for the product of square symmetric matrices, and applied it to a ``multiplicative'' noise model in which each observation of high-dimensional data is drawn independently from some unknown, potentially extensive-rank, covariance matrix \cite{Bun2016}. In that context, knowledge of the overlaps of the data covariance with the unobserved population covariance is sufficient to enable the construction of an optimal rotationally invariant estimator \cite{Ledoit2011,Bun2016, Ledoit2020, potters_bouchaud_2020}.

We have derived analogous results for signals with additive noise: we have computed an asymptotically exact subordination relation for the block resolvent of the free sum of rectangular matrices, i.e. for the resolvent of the Hermitianization of the sum in terms of the resolvents of the Hermitianization of the summands. From the subordination relation, we derived the expected overlap between singular vectors of the sum and singular vectors of the summands. These overlaps quantify how singular vectors are deformed by additive noise. We have calculated separate optimal non-linear singular-value shrinkage expressions for signal denoising and for covariance estimation. Under the assumption of i.i.d. Gaussian noise these shrinkage functions depend only on the noise variance and the empirical data singular value density, which we have shown can be estimated by kernel methods.

We have applied our results in order to study the extensive spike model. We found a significant improvement in estimating signals with even fairly low rank-ratios, over methods that are based on the finite-rank theory. Our results may have significant impact on ongoing research questions around spiked matrix models \cite{el-alaoui18a, Barbier2020, Aubin2021, Ke2021}, such as the question of the detectability of spikes or optimal estimates for the number of spikes, for example.

The subordination relation derived here is closely related to operator-valued free probability, which provides a systematic calculus for block matrices with orthogonally/unitarily invariant blocks, such as the $2\times 2$-block Hermitianizations $\bm{Y},\bm{X},\bm{R}$. In that approach, spectral properties of a matrix are encoded via $2\times 2$ operator-valued Stieltjes and $\xr$ transforms - whose diagonal elements correspond exactly to the block Stieltjes and $\xr$-transforms defined here. A fundamental result in this context is an additive subordination relation for the operator-valued Stieltjes transform, which is an identical formula to \eqref{eq-subordination-stieltjes} \cite{Mingo2017}.

We comment briefly on our derivation of the block resolvent subordination, which is summarized in Section \ref{subsec:subordination} and treated fully in Appendix \ref{sec:Annealed-Appendix}. First, we note that previous work derived resolvent subordination relations for square symmetric matrices using the replica method \cite{Bun2016,Bun2017,potters_bouchaud_2020}. These works assume the replicas decouple which results in a calculation that is equivalent to computing the annealed free energy. Here we used concentration of measure arguments to prove that the annealed approximation is asymptotically correct (Appendix \ref{sec:Justify-Annealed-Appendix}).

In the course of our derivation of the subordination relation we encountered the expectation over arbitrary block-orthogonal rotations of the Hermitianization of the noise matrix (eq \eqref{eq-HCIZ-main} in the main text, and Appendix \ref{sec:rank1-HCIZ-appendix}), which we called a ``block spherical integral''. As noted in the main text, this integral plays an analogous role to the HCIZ spherical integral which appears in the derivation of the subordination relation of square symmetric matrices \cite{potters_bouchaud_2020}. In that setting, the logarithm of the rank-$1$ spherical integral yields the antiderivative of the standard R-transform for square symmetric matrices \cite{Collins2003_HCIZ_Rtransform}. To our knowledge, the particular block spherical integral in our work (Appendix \ref{sec:rank1-HCIZ-appendix}) has not been studied previously. In fact, it is very closely related to the rectangular spherical integral, whose logarithm is the antiderivative of the so-called {\it rectangular} $\xr$-transform \cite{Benaych-Georges2009}. In our setting, two such rectangular spherical integrals are coupled, and the logarithm of the result is the antiderivative of the {\it block} $\xr$-transform \eqref{eq-R-transform} (up to component-wise proportionality constants related to the aspect ratio). While the {\it rectangular} $\xr$-transform is additive, its relationship to familiar RMT objects such as the Stieltjes transform is quite involved. In contrast, the block $\xr$-transform that arises from the block spherical integral is a natural extension of the scalar $\xr$-transform, with a simple definition in terms of the functional inverse of the block Stieltjes transform. Furthermore, as mentioned above, the block $\xr$-transform is essentially a form of the more general operator $\xr$-transform from operator-valued free probability. This formulation is appealing because it provides a direct link between a new class of spherical integrals and operator-valued free probability.

We stress that even under the assumption of Gaussian i.i.d. noise, the optimal estimators we obtained in \ref{sec:optimal rie} are not quite \textit{bona fide} empirical estimators, as they depend on an estimate of the noise variance. This may not be a large obstacle, but we leave it for future work. We do note that while under the assumption of finite-rank signals, appropriate noise estimates can be obtained straightforwardly for example from the median data singular value (see \cite{Gavish2014} for example), this is no longer the case in the extensive regime that we study. In empirical contexts in which one has access to multiple noisy instantiations of the same underlying signal, however, a robust estimate of the noise variance may be readily available.

Other recent work has also studied estimation problems in the extensive-rank regime. \cite{Fleig2022} studied the distribution of pairwise correlations in the extensive regime. \cite{Troiani2022} studied optimal denoising under a known, factorized extensive-rank prior, and arrived at the same shrinkage function we find for the special case of Gaussian i.i.d. noise \eqref{eq-gaussian-shrinkage}. References \cite{Maillard2022} and \cite{Barbier2022} studied both denoising and matrix factorization (dictionary learning) with known, extensive-rank prior.

Lastly, during the writing of this manuscript, the pre-print \cite{Pourkamali2023} presented work partially overlapping with ours. They derived the subordination relation for the resolvent of Hermitianizations as well as the optimal rotationally invariant data denoiser, and additionally establish a relationship between the rectangular spherical integral and the asymptotic mutual information between data and signal. However, unlike our work, this contemporaneous work: (1) does not calculate the optimal estimator of the signal covariance; (2) does not explore the phase diagram of extensive spike model and its associated conceptual insights about the decoupling of singular value phases from singular vector detectability that occurs at extensive but not finite rank; (3) does not extensively numerically explore the inaccuracy and inferior data-denoising and signal-estimation performance of the finite-rank model compared to the extensive-rank model, a key motivation for extensive rank theory; (4) at a technical level \cite{Pourkamali2023} follows the approach of \cite{Bun2016} using a decoupled replica approach yielding an annealed approximation, whereas we prove the annealed approximation is accurate using results from concentration of measure; (5) also at a technical level \cite{Pourkamali2023} employs the rectangular spherical integral resulting in rectangular $\xr$-transforms, whereas we introduce the block spherical integral yielding the block $\xr$-transform, thereby allowing us to obtain simpler formulas. 

We close by noting that our results for optimal estimators depend on the assumption of rotational (orthogonal) invariance. Extending this work to derive estimators for extensive-rank signals with structured priors is an important topic for future study. The rectangular subordination relation and the resulting formulas for singular vector distortion due to additive noise hold for arbitrary signal matrices. These may prove to be of fundamental importance from the perspective of signal estimation in the regime of high-dimensional statistics, as any attempt to estimate the structure of a signal in the presence of noise must overcome both the distortion of the signal's singular value spectrum \textit{and} the deformation of the signal's singular vectors.

\begin{acknowledgments}
  We thank Javan Tahir for careful reading of this manuscript that led to significant improvements. We thank Haim Sompolinsky, Gianluigi Mongillo, Michael Feldman, and Pierre Mergny for helpful comments. S.G. thanks the Simons Foundation and an NSF CAREER award for funding. I.D.L. thanks the Koret Foundation for funding. G.C.M. thanks the Stanford Neurosciences Graduate Program and the Simons Foundation. 
\end{acknowledgments}

\clearpage
\appendix

\section{Finite-Rank Theory for the Spiked Matrix Model}\label{sec:Finite-Spike-Appendix}
We review formulas from \cite{Benaych-Georges2012} for the finite-rank spiked matrix model, $R=sU_1U_2^T+X$, where the $U_a$ are $N_a\times K$ with orthonormal columns, and $X$ is a random $N_1\times N_2$ matrix with well-defined singular value spectrum in the large size limit with fixed aspect ratio $c=\nicefrac{N_1}{N_2}$.

In the case where the noise $X$ is i.i.d. Gaussian with variance $\nicefrac{1}{N_2}$, the critical signal strength below which the signal is undetectable is $s_{crit}=c^{\nicefrac{1}{4}}$. The top $K$ singular values of $R$ are given by

\begin{equation}
  \hat{s}_{l\le K}=\begin{cases}
s\sqrt{\left(1+\frac{c}{s^{2}}\right)\left(1+\frac{1}{s^{2}}\right)} & \mathrm{for}\;s>s_{crit}\\
1+\sqrt{c} & \mathrm{otherwise}.
\end{cases}
\end{equation}
For the square-symmetric setting, see also \cite{Bai2008_finiteK_fluctuations} for derivation of the fluctuations around this asymptotic limit which take the form of the eigenvalues of a $K\times K$ random matrix.

The overlaps of the corresponding singular vectors, $\hat{\bm{u}}_{\bm{a}l}$ for $l=1...K$, with the signal subspaces, $U_a$, for $a=1,2$ are given by 
\begin{align}
  \norm{\hat{\mb{u}}_{\bm{1}l}^TU_1}^{2} & =\begin{cases}
\frac{s^{4}-c}{s^{4}+cs^{2}} & \mathrm{for}\;s>s_{crit}\\
0 & \mathrm{otherwise} \end{cases}\\
\norm{\hat{\mb{u}}_{\bm{2}l}^TU_2}^{2} & =\begin{cases}
\frac{s^{4}-c}{s^{4}+s^{2}} & \mathrm{for}\;s>s_{crit}\\
0 & \mathrm{otherwise} \end{cases}.
\end{align}

For a generic noise matrix, $X$, with block Stieltjes transform, $\bm g^X\lr{\bm{z}}$, \cite{Benaych-Georges2012} defines the $D$-transform, which in our notation is the product of the elements of $\bm g^X\lr{z}$ (where the argument has $z_1=z_2=z$):
\begin{equation}
  D_X\lr{z} = {g}_1^X\lr{z}{g}_2^X\lr{z}
\end{equation}
Then the critical signal satisfies
\begin{equation}
   D_X\lr{x_+}=\frac{1}{s_{crit}^2},
\end{equation}
where $x_+$ is the supremum of the support of the singular value spectrum of $X$.

For suprathreshold signals, $s>s_{crit}$, the data singular value outlier, $\hat s$ satisfies
\begin{equation}
  D_X\lr{\hat s} = \frac{1}{s^2}
\end{equation}
and the two overlaps, corresponding to blocks $a=1$ and $a=2$ for left and right singular vectors, respectively, are given by
\begin{equation}
  \norm{\hat{\bm{u}}_{\bm a}^TU_a}^2 = \frac{-2g_a^X\lr{\hat s}}{D_X\lr{\hat s}D_X'\lr{\hat s}}
\end{equation}

\section{Derivation of the Block-Resolvent Subordination Relation}\label{sec:Annealed-Appendix}

Here we calculate the asymptotic subordination relation \eqref{eq-subordination-main}, found in Section \ref{subsec:subordination} of the main text, for the block resolvent of the free sum of rectangular matrices $R=Y+O_1XO_2^T$, and $O_a$ Haar-distributed orthogonal matrices of size $N_a$ for $a=1,2$. We write $N=N_1+N_2$ and study the large $N$ limit with fixed aspect ratio $c=\nicefrac{N_1}{N_2}$. For notational ease we introduce the ratio of each block's size to the entire matrix:
\begin{equation}
  \beta_a:=\frac{N_a}{N},
\end{equation}
We begin by writing
\begin{equation}
  \bm{M} := zI - \bm{R}=zI-(\bm{Y}-\bar{\bm{O}}\bm{X}\bar{\bm{O}^T}),
\end{equation}
where $\bar{\bm O}=\mat2{O_1}{0}{0}{O_2}$. Next we define the partition function, $\cZ^R\lr{\bm Y} := \lr{\det \bm M}^{-\nicefrac{1}{2}}$, which we can write as a Gaussian integral:
\begin{equation}
\cZ^{R}\left(\bm{Y}\right)=\int\frac{\rm d\bm v}{\sqrt{2\pi}^N}\exp\left(-\frac{1}{2}\bm{v}^{T}\bm{M}\bm{v}\right).
\end{equation}
We also define a corresponding free energy \begin{equation}
  \cF^{R}\left(\bm{Y}\right):=2\log\cZ^{R}\left(\bm{Y}\right),
\end{equation}
and the desired block resolvent is $\bm G^R\lr{z}=\bm M^{-1}=\frac{\rm d}{\rm d\bm Y}\cF^R\lr{\bm Y}$.

Prior work on the case of square symmetric matrices has employed the replica trick to compute this quenched average \cite{Bun2016,Bun2017,potters_bouchaud_2020,Pourkamali2023}. In our notation, this amounts to approximating $\log \cZ^R=\xlim{n}{0}\frac{\lr{\cZ^R}^n-1}{n}$, and then computing $\eO\lrb{\bm{G}^R\lr{z}}=\xlim{n}{0}\eO\lrb{\lr{\cZ^R}^{n-1}\frac{\rm d\cZ^R}{\rm d \bm{Y}}}$ via $n$ Gaussian integrals. Prior work has assumed that the replicas do not couple, which effectively amounts to computing the annealed average, $\log\eO\lrb{\cZ^R\lr{\bm Y}}$.

Instead, we show in Appendix \ref{sec:Justify-Annealed-Appendix} using concentration inequalities that the annealed calculation is in fact asymptotically exact. In particular, as $N\to\infty$, 
\begin{equation}
  \e_{\bar{\bm{O}}} \left[\frac{2}{N}\log\cZ^{R}\left(\bm{Y}\right)\right] \to \frac{2}{N}\log \e_{\bar{\bm{O}}} \left[\cZ^{R}\left(\bm{Y}\right)\right],
\end{equation}

Writing out the expectation and separating factors that depend on
$\bar{\bm{O}}$, we have
\begin{align}
\e_{\bar{\bm{O}}} \left[\cZ^{R}\left(\bm{Y}\right)\right] &= \int\frac{\rm d\bm v}{\sqrt{2\pi}^N}e^{-\frac{1}{2}\bm{v}^{T}\left(zI-\bm{Y}\right)\bm{v}} \nonumber\\ 
& \times \E_{\bar{\bm{O}}}\left[e^{\frac{1}{2}\bm{v}^{T}\bar{\bm{O}}\bm{X}\bar{\bm{O}}^{T}\bm{v}}\right].
\end{align}

The expectation over $\bar{\bm{O}}$ on the right hand side is a rank-$1$ block spherical integral. In Appendix \ref{sec:rank1-HCIZ-appendix}, we derive an asymptotic expression for the expectation, which depends only on, $\bm{\xr}^X$, the block $\xr$-transform of the noise matrix $X$, and the block-wise norms of the vector, $\bm v$. Introducing the two-element vector, $\bm{t}$ whose $a^{th}$ entry is $\frac{1}{N_{a}}\norm{\bm{v_a}}^2$, we have
\begin{equation}
  \E_{\bar{\bm{O}}}\left[e^{\frac{1}{2}\bm{v}^{T}\bar{\bm{O}}\bm{X}\bar{\bm{O}}^{T}\bm{v}}\right] = \exp\lr{\frac{N}{2}H^X\lr{\bm{t}}},
\end{equation}
where in anticipation of a saddle-point condition below, we write $H^X\lr{\bm{t}}$ as a contour integral within $\mathbb{C}^2$ from $0$ to $\bm t$:
\begin{equation}
  H^X\lr{\bm t}:=\int_{0}^{\bm{t}}d\bm{w}\cdot\left(\bm\beta\odot \bm{\xr}^{X}\left(\bm{w}\right)\right),
\end{equation}
where $\bm\beta=\frac{1}{N}\cvec2{N_a}{N_b}$ and $\odot$ is element-wise product. This gives
\begin{align}
\e_{\bar{\bm{O}}} \left[\cZ^{R}\left(\bm{Y}\right)\right]=&\int\frac{d\bm{v}}{\lr{\sqrt{2\pi}}^N}e^{-\frac{1}{2}\bm{v}^{T}\left(zI-\bm{Y}\right)\bm{v}} \nonumber \\
&\times\exp\lr{\frac{N}{2}H^X\lr{\bm{t}}},
\end{align}
 
In order to decouple $\bm{v}$ from $\bm t$, we introduce integration variables and Fourier expressions for the delta-function constraints $\delta\lr{N_a t_a - \norm{\bm{v}_a}^2}$:
\begin{equation}
  1=\int \rm{d}t_a\int \frac{\rm{d}\hat{t}_a}{4\pi i}\exp\lr{-\frac{1}{2}\hat t_a \lr{N_a t_a - \norm{\bm{v}_a}^2}}.
\end{equation}

We now have
\begin{widetext}

\begin{equation}
  \e_{\bar{\bm{O}}} \lrb{\cZ^{R}\lr{\bm{Y}}} = \int \lr{\prod_{a} \frac{\rm{d}t_a \rm{d}\hat{t}_a}{4\pi i}\ex{-\frac{1}{2}N_at_a\hat{t}_a}} \exp\lr{\frac{N}{2} H^X\lr{\bm t}} 
  \int\frac{d\bm{v}}{\lr{\sqrt{2\pi}}^N}  \exp\lr{-\frac{1}{2} \mb{v}^T \left(z I - {\bar{\bm{T}}} -\bm{Y}\right) \mb{v} }, 
\end{equation}
\end{widetext}
where we have introduced the diagonal $N\times N$ matrix, ${\bar{\bm{T}}}$, which has $\hat{t}_1$ along the first $N_1$ diagonal elements followed by $\hat{t}_2$ along the remaining $N_2$ elements.

The integral over $\bm{v}$ is a Gaussian integral with inverse covariance $\lr{zI - \bar{\bm{T}} - \bm{Y} }$ (which is positive-definite for sufficiently large $z$). Crucially, this covariance is exactly, $\bm{G}^Y\mleft(z-\hat{\bm{t}}\mright)$ the block resolvent of $Y$ with a shifted argument, $z-\hat{\bm{t}}$. Note that the block resolvent, as a function of two complex numbers, has emerged here in our calculation.

The result is the inverse square-root of the determinant:
\begin{align}
  \int \rm{d}\mb{v} & \ex{-\frac{1}{2}\mb{v}^T \left(z I-\bm{Y} - {\bar{\bm{T}}}\right) \mb{v} } \propto \det\lr{zI - \bm{Y} -\bar{\bm{T}} }^{-\frac{1}{2}}.
\end{align}

Thus, ignoring proportionality constants we have
\begin{equation}
    \e_{\bar{\bm{O}}} \lrb{\cZ^{R}\lr{\bm{Y}}} \propto \int \rm{d}\bm{t} \rm{d}\hat{\bm{t}}\exp\lr{\frac{N}{2}P^{X,Y}\lr{\bm t,\hat{\bm t}} }, 
\end{equation}
with
\begin{align}\label{eq-final-potential}
  P^{X,Y}\lr{\bm t,\hat{\bm t}} := &-\beta_1t_1 \hat{t}_1 - \beta_2t_2 \hat{t}_2 + H^X\lr{\bm t} \nonumber\\ 
  &-\frac{1}{N}\log\det\lr{zI-\bm Y-\bar{\bm T}}. 
\end{align}
where, we remind the reader, $\beta_a:=\nicefrac{N_a}{N}$.

We expect this integral to concentrate around its saddle point in the large-size limit. We find that taking the derivative of $P^{X,Y}\lr{\bm t,\hat{\bm t}}$ with respect to $t_a$ gives the following appealing saddle-point condition for $\hat{\bm{t}}$:
\begin{equation}
  \hat{\bm{t}} = \bm\xr^X\mleft(\bm{t}\mright).
\end{equation}

In order to take the derivatives with respect to $\hat{t}_a$, we find it helpful to 
write out $N_2$ singular values $s_m$ of $Y$ (including $N_2-N_1$ zeros when $N_2>N_1$). Then $\lr{zI-\bm Y-\bar{\bm T}}$ decouples into $2\times 2$ matrices of the form $\mat2{z-\hat{t}_1}{-s_m}{-s_m}{z-\hat{t}_2}$, and that allows us to write
\begin{align}  
  \det\lr{zI - \bm{Y} -\bar{\bm{T}} } & = \lr{z-\hat{t}_1}^{\lr{N_1-N_2}} \nonumber \\
  &\times \prod_{m=1}^{N_2}\lrb{\lr{z-\hat{t}_1}\lr{z-\hat{t}_2} - s_m^2}\nonumber.
\end{align}
Then we find that taking the derivative of \eqref{eq-final-potential} gives the final saddle-point condition:
\begin{align}
  t_1 &= \lr{z-\hat{t}_2} g_{YY^T}\lr{\lr{z-\hat{t}_1}\lr{z-\hat{t}_2}} \\
  t_2 &= \lr{z-\hat{t}_1} g_{Y^TY}\lr{\lr{z-\hat{t}_1}\lr{z-\hat{t}_2}}. 
\end{align}
We can write this concisely in vector notation:
\begin{equation}\label{eq-last-saddle-Stieltjes-subordination-replica-appendix}
  \bm{t}^* = \bm{g}^Y\lr{z - \bm{\xr}^X\lr{\bm{t}^*}}.
\end{equation}

Thus, asymptotically, the desired free energy is $\eO\lrb{\cF^R\lr{\bm Y}}=NP^{X,Y}\lr{\bm t^*,\bm{\xr}^X\lr{\bm t^*}}$.

Informally, to derive the matrix subordination relation, we differentiate $\cF^R$, which, from \eqref{eq-final-potential}, yields $\lr{zI-\bar{\bm T} - \bm Y}^{-1}=\bm{G}^Y\lr{z-\bm\xr\lr{\bm t^*}}$. But we argued above that $\frac{\rm d}{\rm d\bm Y}\cF^R\lr{\bm Y}=\bm G^R\lr{z}$, which gives the subordination relation.

More formally, consider a Hermitian test matrix, $A$, with a bounded spectral distribution, and then observe that $\frac{1}{N}\frac{\rm d}{\rm d y} \log\det\lr{\bm M + yA}=\tau\lrb{A\bm M^{-1}}$. Thus, we substitute $\bm Y\rightarrow\bm Y+yA$ into the expression for $\cF^R$ \eqref{eq-final-potential} and differentiate to find
\begin{equation} \label{eq-A-deriv-notsolved}
\xlim{N}{\infty}\tau\left[A\eO\lrb{\bm{G}^{R}\left(z\right)}\right] = \xlim{N}{\infty}\tau\left[A\bm{G}^{Y}\left(z-\bm{\xr}^{X}\left(\bm{t}^*\right)\right)\right],
\end{equation}

Using $A$ proportional to either $\mat2{I_{N_1}}{0}{0}{0}$ or $\mat2{0}{0}{0}{I_{N_2}}$, we can now take the normalized block-wise traces of both sides, yielding
\begin{equation}
\label{eq-Stieltjes-subordination-final}
  \bm g^R\lr{z} = \bm{g}^Y\lr{z-\bm{\xr}^X\lr{\bm{t}^*}}.
\end{equation}
Thus, comparing to \ref{eq-last-saddle-Stieltjes-subordination-replica-appendix} we have $\bm t^*=\bm g^R\lr{z}$, and \ref{eq-Stieltjes-subordination-final} becomes the subordination relation for the block Stieltjes transform. Substituting $\bm t^*=\bm g^R\lr{z}$ into \ref{eq-A-deriv-notsolved}, we obtain the desired resolvent relation
\begin{equation}
  \tau\lrb{A\eO\lrb{\bm{G}^R\lr{z}}} = \tau\lrb{A\bm{G}^Y\lr{z-\bm{\xr}^X\lr{\bm{g}^R\lr{z}}}},
\end{equation}
for all Hermitian test matrices $A$ with bounded spectrum, or as written informally in the main text, $\eO\lrb{\bm{G}^R\lr{z}} = \bm{G}^Y\lr{z-\bm{\xr}^X\lr{\bm{g}^R\lr{z}}}$.

\subsection*{Proof the Annealed Free Energy Asymptotically Equals the Quenched Free Energy}\label{sec:Justify-Annealed-Appendix}

Suppose $A,B$ are hermitian matrices with bounded spectrum. Define the function
\begin{equation}\label{eq-f-def-for-Lipschitz}
f\left(O\right) :=\frac{1}{N}\log\det\left(zI-\left(A+OBO^{T}\right)\right),
\end{equation}
for arbitrary orthogonal $O\in\S\O\left(N\right)$. For sufficiently large $z$, the matrix in the determinant is always positive, and this is a smooth function on $\S\O\left(N\right)$ bounded above and below by constants $c_{\pm}:=\log\left(z\pm\left(\norm{A}_{op}+\norm{B}_{op}\right)\right)$, where $\norm{\cdot}_{op}$ is the operator norm. For such $z$, we prove the following Lipschitz bound below (see Section \ref{sec:lipschitz} for proof):
\begin{equation}
\label{eq-Lipschitz-annealed}
\left|f\left(O_{1}\right)-f\left(O_{2}\right)\right|\le\frac{\mu}{\sqrt{N}}\norm{O_{1}-O_{2}}_{2},
\end{equation}
where $\mu:=\pi\norm{B}_{op}e^{-c_{-}}$ and $\norm{\cdot}_{2}$ is the Euclidean norm $\norm{X}_2=\sqrt{\trace{}{\left[X^{T}X\right]}}$. 

In particular, we will be interested in the case that the orthogonal matrix $O$ is block diagonal with blocks $O_a\in \S\O(N_a)$, and thus $O$ is a member of the product space $\S\O(N_1)\times\S\O(N_2)$ with $N_1+N_2=N$. The group $\S\O\left(N_a\right)$ with Haar measure and Hilbert-Schmidt metric obeys a logarithmic Sobolev inequality with constant $\frac{4}{N_a-2}$, so the product space has Sobolev constant $\max_a \frac{4}{N_a-2}=\frac{4}{\gamma N-2}$, where $\gamma:=\min (\frac{N_1}{N},\frac{N_2}{N})$ (\cite{meckes_2019}, Thms. 5.9, 5.16), and we can apply Theorem 5.5 of \cite{meckes_2019}, yielding
\begin{equation}
\P\left[\left|f\left(O\right)-\eO\lrb{f\left(O\right)}\right|\ge\frac{\mu}{\sqrt{N}}r\right]\le2\exp\lr{-\left(\gamma N-2\right)\frac{r^{2}}{8}},
\end{equation}
for all $r\ge0$.

Writing $H:=\E_{O}\left[f\left(O\right)\right]$, and defining $M:=z-A-OBO^{T}$ (so that $\det M=e^{Nf\lr{O}}$), this implies
\begin{equation}
  \P\left[\det\left(M\right)\ge e^{NH+\sqrt{N}\mu r}\right]\le2e^{-\left(\gamma N-2\right)\frac{r^{2}}{8}}.
\end{equation}
Since $\det\left(M\right)\le e^{Nc_{+}}$, we can upper bound the expectation:
\begin{multline}
\E_{O}\left[\det\left(M\right)\right]\le\left(1-2e^{-\left(\gamma N-2\right)\frac{r^{2}}{8}}\right)e^{NH+\sqrt{N}\frac{\pi\mu}{2}r}\\ 
+2e^{-\left(\gamma N-2\right)\frac{r^{2}}{8}}e^{Nc_{+}}.
\end{multline}
Choosing $r=\sqrt{8c_{+}/\gamma}$, we find that $\frac{1}{N}\log\E_{O}\left[\det\left(M\right)\right]$ is less than or equal to
\begin{multline*}
 \frac{1}{N}\log\left(\left(1-2e^{-\left(N-2\right)c^{+}/\gamma}\right)e^{NH+\sqrt{N}\frac{\pi\mu}{2}r}+2e^{2c_{+}\gamma}\right)\\
\xrightarrow{N\to\infty}H=\frac{1}{N}\E_{O}\left[\log\det\left(M\right)\right],
\end{multline*}
which shows that the limiting annealed average is less than or equal to the limiting quenched average. We could obtain a lower bound via a similar argument, but we have directly via Jensen's inequality that the quenched average
is less than or equal to the annealed average, $\frac{1}{N}\E_{O}\left[\log\det\left(M\right)\right]\le\frac{1}{N}\log\E_{O}\left[\det\left(M\right)\right]$,
so in the limit they are equal:
\begin{equation}
\lim_{N\to\infty}\frac{1}{N}\E_{O}\left[\log\det\left(M\right)\right]=\lim_{N\to\infty}\frac{1}{N}\log\E_{O}\left[\det\left(M\right)\right].
\end{equation}

\subsubsection{Lipschitz Bound}\label{sec:lipschitz}

To prove \ref{eq-Lipschitz-annealed}, note that the gradient of $f$ \eqref{eq-f-def-for-Lipschitz} is
\begin{equation}
  \nabla_{O}f\left(O\right)=-2\frac{1}{N}M^{-1}OB,
\end{equation}
where, as above, $M=z-\left(A+OBO^{T}\right)$. Thus, the ordinary Euclidean norm of the gradient is
\begin{align}
\norm{\nabla_{O}f\left(O\right)}_2 & =2\frac{1}{N}\norm{M^{-1}OB}_2\\
 & =\frac{2}{N}\sqrt{\trace{}{\left[M^{-2}OB^{2}O^{T}\right]}}.
\end{align}
$M^{-2}$ and $OB^{2}O^{T}$ are positive definite Hermitian matrices, so $\trace{}{\left[M^{-2}OB^{2}O^{T}\right]} \le N\norm{B^{2}}_{op}\norm{M^{-2}}_{op}$. From $M$'s definition we have $\norm{M^{-2}}_{op}\le\left(z-\left(\norm{A}_{op}+\norm{B}_{op}\right)\right)^{-2}=e^{-2c_{-}}$,
and so
\begin{equation}
\norm{\nabla_{O}f\left(O\right)}_2\le\frac{2\norm{B}_{op}}{e^{c_{-}}\sqrt{N}}.
\end{equation}
This shows that $f$ changes by at most $\frac{2\norm{B}_{op}}{e^{c_{-}}\sqrt{N}}$ times the geodesic distance on the group: $\left|f\left(O_{1}\right)-f\left(O_{2}\right)\right|\le\frac{2\norm{B}_{op}}{e^{c_{-}}\sqrt{N}}d\left(O_{1},O_{2}\right)_{\S\O\left(N\right)}$.
The geodesic distance is upper bounded by $\pi/2$ times the Euclidean distance (\cite{meckes_2019},
p159), so 
\begin{equation}
\left|f\left(O_{1}\right)-f\left(O_{2}\right)\right|\le\frac{\pi\norm{B}_{op}}{e^{c_{-}}\sqrt{N}}\norm{O_{1}-O_{2}}_{2}.
\end{equation}

\section{Rank-$1$ Block Spherical HCIZ Integral}\label{sec:rank1-HCIZ-appendix}

In this section we introduce the ``block spherical integral'', which extends the HCIZ integral to the setting of Hermitianizations of rectangular matrices.

We consider an $N_1\times N_2$ matrix, $X$, with $N=N_1+N_2$ and consider the limit of large $N$ with fixed $c=\nicefrac{N_1}{N_2}$. For notational ease we will introduce
\begin{equation}
  \beta_a :=\frac{N_a}{N},
\end{equation}
for both $a=1,2$.

In the general-rank setting we write
\begin{equation}
  I^X\lr{\bm T} := \eO\lrb{\exp\lr{\frac{N}{2} \rm{Tr}\bm{T} \bar{\bm O}\bm X\bar{\bm O}^T}},
\end{equation}
where $\bar{\bm O}=\mat2{O_1}{0}{0}{O_2}$ is a block-orthogonal matrix, i.e. both $O_a$ are Haar distributed $N_a\times N_a$ matrices, and $\bm T$ is an arbitrary $N\times N$ matrix.

We here solve the rank-$1$ case, which arises in Appendix \ref{sec:Annealed-Appendix} in the calculation of the subordination relation. In order to match the normalization there, we write $T=\frac{1}{N}\bm{v}\bm{v}^T$, where the individual elements, $v_i$, are $O\lr{1}$. We have
\begin{equation}
  I^X\lr{\bm T} := \eO\lrb{\exp\lr{\frac{1}{2} \bm{v}^T \bar{\bm O}\bm X\bar{\bm O}^T \bm {v}}}.
\end{equation}

We write $\bm v=\cvec2{\bm{v_1}}{\bm{v_2}}$ in block form, and observe that the block-orthogonal $\bar{\bm O}$ preserves the within-block norms of $\bm v$. Therefore, we define 
\begin{equation}
  \bm{w}_a := O_a^T \bm{v}_a,
\end{equation}
for $a=1,2$, and perform integrals over arbitrary $\bm{w}=\cvec2{\bm{w_1}}{\bm{w_2}}$ while enforcing norm-constraints within blocks. We define the $2$-component vector $\bm t$:
\begin{equation}\label{eq-blockwise-norm-rank-1-HCIZ}
  t_a^ := \frac{1}{N_a} \norm{\bm{v}_a}^2
\end{equation}

Then we can write:
\begin{equation}\label{eq-1d-HCIZ-rank1-appendix}
  I^X\lr{\bm{T}}=\frac{Z\lr{\bm{t},X}}{Z\lr{\bm{t},0}},
\end{equation}
where we have defined 
\begin{align}
  Z\lr{\bm{t},X}:=&\int \frac{\rm{d}\bm{w}}{\lr{2\pi}^{\nicefrac{N}{2}}} 
\exp\lr{\frac{1}{2}\bm{w}^T\bm X\bm{w}} \nonumber\\
&\times\prod_{a=1,2}\delta\lr{\norm{\bm{w}_a}^2-N_a t_a}.
\end{align}
We calculate $Z\lr{\bm{t},X}$ by using the Fourier representation of the delta function, over the imaginary axis: $\delta\lr{x}=\int_{-i\infty}^{+i\infty}\frac{\exp\lr{-\nicefrac{qx}{2}}}{4\pi i}\rm{d}q$. This gives

\begin{align}
  Z\lr{\bm{t},X}:=&\int \lr{\prod_{a=1,2}\frac{\rm{d}q_a}{4\pi i} \ex{\frac{1}{2}\lr{N_aq_at_a}}} \\
  &\int \frac{\rm{d}\bm{w}}{\lr{2\pi}^{\nicefrac{N}{2}}}\ex{-\frac{1}{2} \bm{w}^T\lr{\bar{\bm{Q}}-\bm{X}}\bm{w}}\nonumber
\end{align}
where we have introduced the $N\times N$ diagonal matrix, $\bar{\bm Q}$, which has $q_1$ on its first $N_1$ diagonal elements, and $q_2$ on the remaining $N_2$ elements.

The Gaussian integral over $\bm{w}$ now yields $\det\lr{\bar{\bm{Q}}-\bm{X}}^{-\nicefrac{1}{2}}$. Writing $N_2$ singular values of $X$ as $x_{m}$ (which includes $N_2-N_1$ zeros in the case $N_2>N_1$), we can write:

\begin{equation}
  \det\lr{\bar{\bm{Q}}-\bm{X}} = q_1^{\lr{N_1-N_2}}\prod_{m=1}^{N_2}\lr{q_1q_2 - x_m^2}
\end{equation}

Thus, at this stage we have

\begin{equation}\label{eq-1st-saddle-pt-HCIZ-rank1-appendix}
  Z\lr{\bm{t},X}:=\int_{-i\infty}^{i\infty} \frac{\rm{d}q_1\rm{d}q_2}{\lr{4\pi i}^2} \exp\lrb{\frac{N}{2} F^X\lr{\bm{t},\bm{q}}}
\end{equation}
with
\begin{align}
  F^X\lr{\bm{t},\bm{q}} =& \beta_1q_1t_1 + \beta_2q_2t_2 +\lr{\beta_2-\beta_1}\log q_1\nonumber\\
  &- \frac{1}{N}\sum_{m=1}^{N_2}\log\lr{q_1q_2 - x_m^2}
\end{align}

To find the saddle-point, we take partial derivatives with respect to $q_1$ and $q_2$, and find:
\begin{align}
  t_1 &= q_2^* g_{XX^T}\lr{q_1^* q_2^*} \\
  t_2 &= q_1^* g_{X^TX}\lr{q_1^* q_2^*},
\end{align}

For notational clarity, in this section we define the functional inverse of the block Stieltjes transform, $\bm\xb^X\lr{\bm{t}} := \lr{\bm{g}^X}^{-1}\lr{\bm t}$, satisfying
\begin{equation}
  \bm\xb^X\lr{\bm g^X\lr{\bm z}} = \bm{z}.
\end{equation}
In the limit of large $z_1,z_2$, we have $g^X_1\lr{\bm{z}}\approx \frac{1}{z_2}$ and $g^X_2\lr{\bm{z}}\approx\frac{1}{z_1}$, and therefore for small $t_1,t_2$ we have $\xb^X_1\lr{\bm{t}}\approx \frac{1}{t_2}$ and $\xb^X_2\lr{\bm{t}}\approx \frac{1}{t_1}$. Generally, the functional inverse, $\bm{\xb}^X\lr{\bm{t}}$ exists for $\bm{t}$ with sufficiently small norm.

Thus, the saddle-point condition for $Z\lr{\bm t, \bm X}$ can be written succinctly as $\bm q^* = \bm{\xb}^X\lr{\bm t}$.

Finally, we find the asymptotic value of $I^X\lr{\bm v}$ \eqref{eq-1st-saddle-pt-HCIZ-rank1-appendix} by also solving the saddle-point for $Z\lr{\bm t,0}$. For $X=0$ we have $g_{XX^T}{z}=z^{-1}$, so that the saddle-point condition for $Z\lr{\bm t,0}$ is simply $q_a^*=t_a^{-1}$. This yields $F^0\lr{\bm t, \bm q^*} = \sum_a\beta_a\lr{1+\log t_a}$

We therefore arrive at our asymptotic approximation for the rank-$1$ block spherical integral:
\begin{equation}
  I^X\lr{\mb T} = \exp{\frac{N}{2}H^X\left(\bm t\right)},
\end{equation}
where we have:
\begin{align}
  H^X\lr{\mb t} =& \sum_{a=1,2}\beta_a\lr{t_a\xb_a^X\lr{\mb t}-\log t_a -1} \nonumber \\
  & - \frac{1}{N}\log\det\lr{\bar{\bm{\xb}}^X\lr{\bm t} - \bm X}
\end{align}
where we have written $\bar{\bm \xb}^X\lr{\bm t}$ to indicate the $N\times N$ diagonal matrix with $\xb_1^X\lr{\bm t}$ along the top $N_1$ diagonal elements, and $\xb_2^X\lr{\bm t}$ along the remaining $N_2$. 

We observe an appealing relationship between the rank-$1$ block spherical integral and the block $\xr$-transform. By the saddle-point conditions, the partial derivatives of $F^X\lr{\bm t,\bm{q}^*}$ with respect to $q_a$ are zero. Therefore the gradient of $H^X$ with respect to $\bm t$ treats $\bm{\xb}^X$ as constant, and we have simply

\begin{equation}
  \frac{\rm d H^X\lr{\bm t}}{\rm{d} t_a} = \beta_a\lr{\xb_a^X\lr{\bm t} - \frac{1}{t_a}} = \beta_a\xr_a^X\lr{\bm t}.
\end{equation}\label{eq-R-transform-rank1-appendix}

We therefore write $H^X\lr{\bm t}$ as a contour integral in $\mathbb C^2$:
\begin{equation}
  H^X\lr{\bm t} = \int_0^{\bm t} \rm d\bm w\cdot \lr{\bm{\beta}\odot \bm{\xr}^X\lr{\bm t}},
\end{equation}
where $\odot$ is the element-wise product and $\bm{\beta}=\cvec2{\beta_1}{\beta_2}=\frac{1}{N}\cvec2{N_1}{N_2}=\frac{1}{1+c}\cvec2{c}{1}$.

\section{The Block $\xr$-Transform of Gaussian Noise}\label{sec:Gaussian-noise-Appendix}

In this section we calculate the block $\xr$-transform for the $N_1\times N_2$ (with $c=\nicefrac{N_1}{N_2}$) matrix $X$ with i.i.d. Gaussian elements: $X_{ij}\sim \xnorm\lr{0,\frac{\sigma^2}{N_2}}$.

For notational clarity, in this section we write the functional inverse of the block Stieltjes transform for any rectangular matrix, $A$, as $\bm\xb^A\lr{\bm{t}} := \lr{\bm{g}^A}^{-1}\lr{\bm t}$, satisfying
\begin{equation}
  \bm\xb^A\lr{\bm g^A\lr{\bm z}} = \bm{z}
\end{equation}

Note that from the definition of $\bm{g}^A\lr{\bm z}$ one can find a relationship between the two elements of the inverse block Stieltjes transform:

\begin{equation}\label{eq-inverse-Stieltjes-components}
t_2\xb_2^A\lr{\bm{t}}=ct_1\xb_1^A\lr{\bm{t}},
\end{equation}
where $c$ is the aspect ratio of $A$.

The block $\xr$-transform is defined as
\begin{equation}
  \bm{\xr}^A\lr{\bm{t}} = \bm\xb^A\lr{\bm t} - \frac{1}{\bm{t}}
\end{equation}
where the multiplicative inverse, $\nicefrac{1}{\bm{t}}$, is element-wise.

To find $\bm\xb^X$, we observe that in general the product of the two elements of $\bm {g}^X\lr{\bm z}$ is a scalar function that depends only on the product of the elements of $\bm z$. That is, $g_1^X\lr{\bm z}g_2^X\lr{\bm z} = z_1z_2g_{XX^T}\lr{z_1z_2}g_{X^TX}\lr{z_1z_2}$. Therefore, we define
\begin{equation}
 {\Lambda}_X\lr{z}:= zg_{XX^T}\lr{z}g_{X^TX}\lr{z}
\end{equation}

We can find $\bm\xb^X\lr{\bm t}$ by first inverting $\Lambda_X\lr{z}$, and then
\begin{equation}
  \xb_1^X \lr{\bm{t}}\xb_2^X\lr{\bm{t}} = {\Lambda}_X^{-1}\lr{t_1 t_2}
\end{equation}

For the Gaussian matrix, $X$, we have
\begin{equation}
  g_{XX^T}\lr{z} = \frac{z+\sigma^2\lr{1+c} - \sqrt{\lr{z-x_+^2}\lr{z-x_-^2}}}{2z\sigma^2}
\end{equation}
with
\begin{equation}
  x_{\pm} = \sigma \lr{1\pm\sqrt{c}}
\end{equation}

From there we find
\begin{equation}
  \Lambda_X\lr{z}=\frac{z-\sigma^2\lr{1+c} + \sqrt{\lr{z-x_+^2}\lr{z-x_-^2}}}{2c\sigma^4}
\end{equation}
Some further algebra yields
\begin{equation}
  \Lambda_X^{-1}\lr{t} = c\sigma^4t + \frac{1}{t} + \sigma^2\lr{1+c}
\end{equation}
Thus, we have
\begin{equation}\label{eq-intermediate-R-transform-step}
  \xb_1^X\lr{\bm{t}} \xb_2^X\lr{\bm{t}} = c\sigma^4t_1 t_2 + \frac{1}{t_1 t_2} + \sigma^2\lr{1+c}
\end{equation}
Using the general relationship between elements of $\bm{\xb}\lr{\bm{t}}$ \eqref{eq-inverse-Stieltjes-components} yields a quadratic equation for $\xb_1^X$. We choose the root that yields $\xb_1^X\approx \nicefrac{1}{t_1}$ in the large $\bm{t}$ limit, and then use \eqref{eq-inverse-Stieltjes-components} to find $\xb_2^X$,  finally arriving at
\begin{align}
  \xb^X_1\lr{\bm t} & =\frac{1}{t_1} + \sigma^2 t_2 \\
  \xb^X_2\lr{\bm t} &= \frac{1}{t_2} + c\sigma^2 t_1
\end{align}

Finally this yields for the block R-transform:
\begin{equation}
  \bm\xr^X\lr{\bm t} = \sigma^2\cvec2{t_2}{c t_1}
\end{equation}

As a side note we point out that from the relationship between elements of the block Stieltjes inverse \eqref{eq-inverse-Stieltjes-components} it follows that 
\begin{equation}
  t_2\xr^A_2\lr{\bm t}=ct_1\xr^A_1\lr{\bm t},
\end{equation}
for all rectangular $A$ with aspect ratio $c$.

\section{Block Stieltjes Transform and Singular Value Density of the Extensive Spiked Model}\label{sec:Stieltjes-polynomial-Appendix}
We report the quartic equation from Section \ref{subsec:SVD of extensive spike} for the first element of the block Stieltjes transform of the $N_1\times N_2$ extensive spiked model, $R=sU_1U_2+X$, with rank-ratio $b=\nicefrac{K}{N_1}$ and aspect ratio $c=\nicefrac{N_1}{N_2}$. For notational simplicity, in this section we write $g:=g_1^R\lr{z}$. Multiplying out \eqref{eq-extensive-spike-Stieltjes-polynomial} yields a quartic:
\begin{equation}
  Ag^4 + Bg^3 + Cg^2 + Dg + E = 0
\end{equation}
with
\begin{align}
  A &= -c^3 \\
  B &=c^2\lr{3z - 2\frac{1-c}{z}} \\
  C &= c\lr{-3z^2 - \lr{\frac{1-c}{z}}^2 - 5c + 4+s^2} \\
  D &= z^3 + \lr{4c-2-s^2}z + \lr{1-2c+s^2}\frac{1-c}{z}\\
  E &= -z^2 -c +1 +\lr{1-b}s^2
\end{align}
In order to obtain the singular value density $\rho_1^R\lr{\hat s}$ numerically, we use the roots method of the NumPy polynomial class in Python (3.9.7), to solve with $z=\hat s - 10^{-7}i$, and select the root with the largest imaginary part. To our knowledge there is no guarantee that the root with largest imaginary part is the correct root, but we find this works in practice.

\section{Kernel Estimates of Empirical Spectral Densities}\label{sec:Kernel-estimates-empirical-Appendix}
In order to employ our optimal estimators in empirical settings (Section \ref{sec:optimal rie}), we need to be able to estimate the block Stieltjes transform, $g_1^R\lr{\hat s}$, from data. Developing optimal algorithms to achieve this is left for future work, but here we use a technique inspired by \cite{potters_bouchaud_2020} and \cite{Ledoit2020} to demonstrate proof of principle. We use this kernel on the extensive spike model in Figure \ref{fig:Figure_8_Optimal_Estimation}C.

Given $N_1$ data singular values, $\{\hat{s}_m\}$ (assuming $N_1<N_2$ without loss of generality), we define a smoothed block Stieltjes transform:
\begin{equation}
\tilde{g}_1^R\lr{z} := \frac{1}{N_1}\sum_{m=1}^{N_1}\frac{z}{z^2 - \hat{s}_m^2 - i\eta_m}
\end{equation}
where $\eta_m$ is a local bandwidth term given by:
\begin{equation}
  \eta_m = \frac{\hat s_m}{N^{\nicefrac{1}{2}}}.
\end{equation}

\bibliography{Literature}

\end{document}